\documentclass{aa}  

\usepackage{graphicx}
\usepackage{txfonts}
\usepackage{xcolor}

\begin{document}

   \title{Nuclear star cluster formation in star-forming dwarf galaxies\thanks{Based on observations collected at the ESO Paranal La Silla Observatory, Chile, Prog. ID 108.220K.001, 0104.D-0503, 0100.B-0116, and 095.B-0.532.}}

   \subtitle{}

   \author{Katja Fahrion\thanks{ESA research fellow}\fnmsep
          \inst{1}
          \and
          Teodora-Elena Bulichi
          \inst{1}\fnmsep\inst{2}
          \and
          Michael Hilker
          \inst{3}
          \and
          Ryan Leaman
          \inst{4}
          \and
          Mariya Lyubenova
          \inst{3}
          \and
          Oliver M\"uller
          \inst{5}
          \and
          Nadine Neumayer
          \inst{6}
          \and
          Francesca Pinna
          \inst{6}
          \and
          Marina Rejkuba
          \inst{3}
          \and
          Glenn van de Ven
          \inst{4}
    }

   \institute{European Space Agency (ESA), European Space Exploration and Research Centre (ESTEC), Keplerlaan 1, 2201 AZ Noordwijk, the Netherlands.
             \email{katja.fahrion@esa.int}
             \and
            Institute for Astronomy, University of Edinburgh, Royal Observatory, Blackford Hill, Edinburgh.
        EH9 3HJ, United Kingdom
            \and 
        European Southern Observatory, Karl-Schwarzschild-Strasse 2, 85748 Garching bei M\"{u}nchen, Germany
        \and
        Department of Astrophysics, University of Vienna, T\"urkenschanzstrasse 17, 1180 Wien, Austria
        \and
        Institute of Physics, Laboratory of Astrophysics, Ecole Polytechnique F\'{e}d\'{e}rale de Lausanne (EPFL), 1290 Sauverny, Switzerland
        \and
        Max-Planck Institute for Astronomy, K\"{o}nigstuhl 17, 69117 Heidelberg, Germany
             }

   \date{\today}

 
  \abstract
   {Nuclear star clusters (NSCs) are massive star clusters found in all types of galaxies from dwarfs to massive galaxies. Recent studies show that while low-mass NSCs in dwarf galaxies ($M_\text{gal} < 10^{9} M_\odot$) form predominantly out of the merger of globular clusters (GCs), high-mass NSCs in massive galaxies have assembled most of their mass through central enriched star formation. So far, these results of a transition in the dominant NSC formation channel have been based on studies of early-type galaxies and massive late-type galaxies. Here, we present the first spectroscopic analysis of a sample of nine nucleated late-type dwarf galaxies with the aim of identifying the dominant NSC formation pathway. We use integral-field spectroscopy data obtained with the Multi Unit Spectroscopic Explorer (MUSE) instrument to analyse the ages, metallicities, star formation histories, and star formation rates of the NSCs and their surroundings. Our sample includes galaxies with stellar masses $M_\text{gal} = 10^7 - 10^9 M_\odot$ and NSC masses $M_\text{NSC} = 6 \times 10^4 - 6 \times 10^{6} M_\odot$. Although all NSC spectra show emission lines, this emission is not always connected to star formation within the NSC, but rather to other regions along the line of sight. The NSC star formation histories reveal that metal-poor and old populations dominate the stellar populations in five NSCs, possibly stemming from the inspiral of GCs. The NSCs of the most massive galaxies in our sample show significant contributions from young and enriched populations that indicate additional mass growth through central star formation. Our results support previous findings of a transition in the dominant NSC formation channel with galaxy mass, showing that the NSCs in low-mass galaxies predominantly grow through the inspiral of GCs, while central star formation can contribute to NSC growth in more massive galaxies.}


   \maketitle
%

\section{Introduction}
Nuclear star clusters (NSCs) are massive, compact star clusters found in the centres of a majority of galaxies of all types. With typical masses between 10$^4$ and 10$^8$ $M_\sun$ and effective radii of 1 - 20 pc, they are the densest stellar systems in the Universe (see the review by \citealt{Neumayer2020} and references therein), surpassing the stellar densities found in globular clusters (GCs, see \citealt{Brodie2006} or \citealt{Beasley2020} for reviews on GCs).

Concerning the formation of NSCs, two main pathways are discussed that impart different signatures in the NSC  properties such as internal kinematics, ages, metallicities, and star formation histories (SFHs). In the GC accretion or GC inspiral scenario, the NSC forms out of the (gas-free) merger of GCs that spiral into a galaxy's centre due to dynamical friction (e.g. \citealt{Tremaine1975, CapuzzoDolcetta1993, CapuzzoDolcetta2008, Agarwal2011, Portaluri2013, Antonini2013, ArcaSedda2014, Gnedin2014}). In this scenario, a NSC is expected to reflect typical properties of GCs, which usually are characterised by simple SFHs and low metallicity. In particular, since GCs usually are significantly less enriched than the central regions of their host galaxies (e.g. \citealt{Lamers2017}), the GC accretion scenario can explain the metal-poor NSCs recently found in some dwarf galaxies \citep{Fahrion2020, Johnston2020, Fahrion2021}. We note here that while GCs in massive galaxies such as the Milky Way are very old ($>$ 10 Gyr, e.g. \citealt{Leaman2013}), dwarf galaxies can also host younger and intermediate-age GCs (2 $-$ 10 Gyr, \citealt{Puzia2008, Martocchia2018, Bica2020}).

As an alternative formation path, the in situ star formation scenario considers a NSC to form directly at the galaxy's centre out of gas (e.g. \citealt{Loose1982, Milosavljevic2004, Schinnerer2007, Bekki2006, Bekki2007, Antonini2015}). As this star formation can proceed in several episodes during a galaxy's evolution, a NSC formed in that way might exhibit a complex SFH and young stellar ages. Depending on the metallicity of the accreted gas, the NSC metallicity that can surpass those of typical GCs due to efficient enrichment at the bottom of the galaxy's gravitational potential where metals can be retained. Such signatures of in situ formation have been identified spectroscopically in NSCs of massive early-type hosts ($M_\ast > 10^9 M_\sun$, \citealt{Fahrion2019, Fahrion2021, Pinna2021}) and NSCs located in the centres of massive late-type galaxies \citep{Kacharov2018, Pinna2021, Hannah2021}. 

Additionally, hybrid scenarios have been discussed, for example describing the inspiral of gas-rich young star clusters (several hundred million years) that are formed in the central regions of a galaxy \citep{Guillard2016, Paudel2020} and then spiral into the galaxy centre within a few hundred million years \citep{ArcaSedda2015}. 
Such a formation pathway can lead to similar stellar population properties in the NSC as the in situ formation channel because both include star formation in the central region of a galaxy that involves pre-enriched material, in contrast to the accretion of generally metal-poor GCs. While the in situ channel can lead to higher angular momentum of the formed NSC (e.g. \citealt{Seth2010, Lyubenova2019, Pinna2021}), in unresolved studies it is challenging to distinguish between in situ formation and accretion of young, gas-rich clusters. Nevertheless, both can be understood as NSC formation through central star formation.

Recent studies based on integral-field spectroscopy that enable a comparison of NSC kinematics and stellar population properties to that of the central parts of the host galaxy have been successful in identifying the dominant NSC formation pathway in individual galaxies, both for unresolved NSCs \citep{Fahrion2020, Johnston2020, Fahrion2021} and in studies of spatially resolved NSCs \citep{Lyubenova2019, Pinna2021, Hannah2021}. Using a sample of 25 nucleated early-type galaxies located in the Fornax and Virgo galaxy clusters that were observed with the Multi-Unit Spectroscopic Explorer (MUSE) instrument, \cite{Fahrion2021} identified a clear trend in the dominant NSC formation channel with galaxy and NSC mass. While the low-mass NSCs ($M_\text{NSC} < 10^7 M_\sun$) in early-type galaxies dwarf ($M_\text{gal} < 10^9 M_\sun$) appear to form predominantly via the inspiral of GCs (see also, e.g., \citealt{Johnston2020, Su2022}), massive NSCs in more massive galaxies have assembled most of their mass through central star formation (see also e.g. \citealt{Turner2012, SanchezJanssen2019, Neumayer2020}). At intermediate NSC and galaxy masses, \cite{Fahrion2021} found evidence of both channels contributing to the formation of individual NSCs as evident from both old and metal-poor populations as well as young and metal-rich populations (see also \citealt{Kacharov2018, Johnston2020, Hannah2021}). 

While spectroscopy of NSCs in massive late-type galaxies (LTGs) has found indications of NSCs forming predominantly through central star formation (e.g. \citealt{Kacharov2018, Pinna2021, Hannah2021}), nucleated dwarf LTGs have not yet been explored spectroscopically in large numbers (but see \citealt{Leaman2020}), although they have been studied photometrically (e.g. \citealt{Georgiev2009a, Georgiev2009b, Paudel2020}). In this paper, we build on the work presented in \cite{Fahrion2021} by extending the analysis of NSC formation channels to dwarf LTGs in low-density environments.
We acquired MUSE data for a sample of nine nucleated dwarf LTGs with galaxy masses $\leq 10^9 M_\sun$. We derived ages, metallicities, SFHs and star formation rates (SFRs) of their NSCs in comparison to the host galaxies and use these diagnostics to constrain their NSC formation pathways, differentiating between GC accretion and central star formation. 

The following section introduces the galaxy sample and the MUSE data. Section \ref{sect:methods} describes our methods of extracting the stellar population properties of the NSCs and their surroundings, and Sect. \ref{sect:results} presents our results. We discuss these in Sect. \ref{sect:discussion} and conclude in Sect. \ref{sect:conclusions}. 

\section{Data}
\label{sect:sample}

\begin{figure*}
    \centering
    \includegraphics[width=0.95\textwidth]{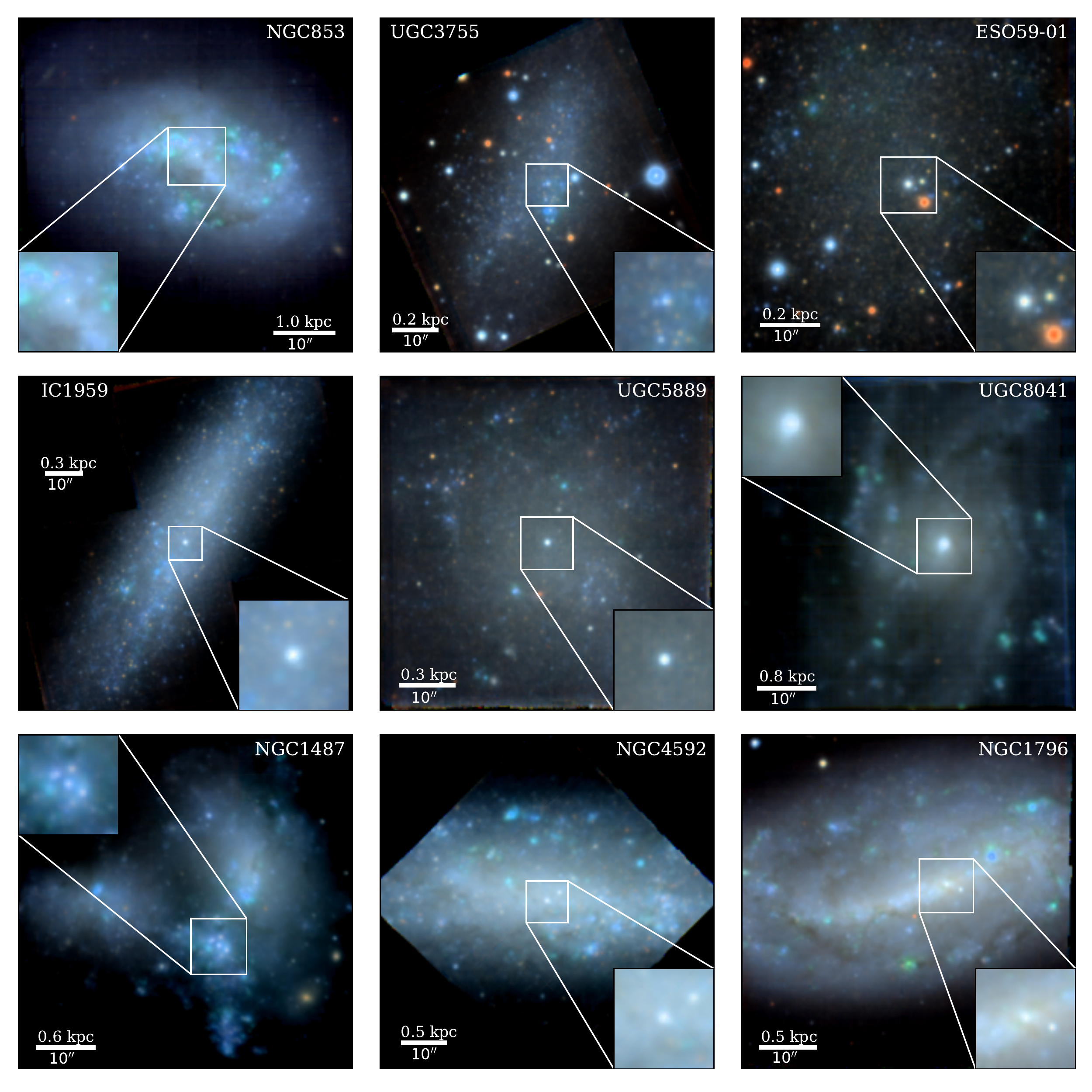}
    \caption{RGB images of the sample galaxies based on synthetic $V$, $I$, and $R$ band images obtained from the MUSE cubes. The inset shows a zoom to a 10\arcsec $\times$ 10\arcsec region around each NSC. The conversion from arcseconds to kpc is shown in white. North is up, east is to the left.}
    \label{fig:galaxy_sample}
\end{figure*}

\begin{table*}[!htb]
\caption{Galaxy properties}\vspace{-4mm}
\begin{center}
\begin{tabular}{c c c c c c c c c c} \hline\hline
Galaxy & RA & DEC & D & E(B-V) & log($M_\text{gal}$) & log($M_\text{NSC}$) & Exp. time & <FWHM> & Ref \\
    &  &  & (Mpc) & (mag) & & & (sec) & arcsec & \\ 
(1) & (2) & (3) & (4) & (5) & (6) & (7) & (8) & (9) & (10) \\    \hline
NGC\,853 & 02:11:41.19 & $-$09:18:21.6 & 21.0 & 0.023 & 6.96 & 6.31 & 2000 & 0.59 & d, e \\
UGC\,3755 & 07:13:51.60 & $+$10:31:19.0 & 4.99 & 0.078 & 7.75 & 4.78 & 4 $\times$ 2000 & 0.57 &  d, e \\
ESO\,059-01 & 07:31:18.20 & $-$68:11:16.8 & 4.57 & 0.147 & 8.1 & 6.16 & 4 $\times$ 2000 &  0.53 & a, b, c\\
IC\,1959 & 03:33:12.59 & $-$50:24:51.3 & 6.05 & 0.011 & 8.2 & 6.13 & 4 $\times$ 2000 & 0.66 & a, b, c  \\
UGC\,5889 & 10:47:22.30 & $+$14:04:10.0 & 6.89 & 0.031 & 8.20 & 6.09 & 4 $\times$ 2000 & 0.58 & d, e \\
UGC\,8041 & 12:55:12.66 & $+$00:06:59.9  & 17.1 & 0.019 & 8.83 & 6.74 & 2405 & 1.24 & d, e \\
NGC\,1487 & 03:55:46.10 & $-$42:22:05.0 & 12.1 & 0.010 & 8.95 & 6.01 & 2000 + 2 $\times$ 3600 & 0.56 &  d, e \\
NGC\,4592 & 12:39:18.74 & $+$00:31:55.2 & 10.6 & 0.042 & 9.02 & 5.80 & 1800 & 0.84 & d, e \\
NGC\,1796 & 05:02:42.55 & $-$61:08:24.2 & 10.6 & 0.021 & 9.06 & 6.82 & 2000 & 0.55 &  d, e \\
 \hline
\end{tabular}
\label{tab:galaxies}
\end{center}
\tablefoot{(1) galaxy name, (2), (3) right ascension and declination, (4) distance in Mpc, (5) foreground extinction, (6, 7) galaxy and NSC stellar masses, (8) exposure time in seconds, (9) average FWHM of the PSF and (10) references: a - \cite{Georgiev2009a}, b - \cite{Georgiev2009b}, c - \cite{Yu2020}, e - \cite{Georgiev2014}, d - \cite{Georgiev2016}.}
\end{table*}

We use integral-field spectroscopy data of nine nucleated dwarf LTGs observed with the MUSE instrument \citep{MUSE} which is mounted at the Very Large Telescope of the European Southern Observatory (ESO) in Chile. MUSE offers a continuous 1\arcmin$\times$1\arcmin\, field of view with a spatial sampling of 0.2\arcsec\,per pixel. Along the wavelength dimension, MUSE covers the optical regime from 4700 to 9300 \AA\,with a spectral sampling of 1.25\,\AA\,per pixel and an average instrumental resolution of $\sim 2.5\,\AA$. 

\subsection{Data reduction}
Seven of the nine galaxies used here (ESO\,59-01, IC\,1959, NGC\,1487, NGC\,1796, NGC\,853, UGC\,3755, and UGC\,5889) were observed in between September 2021 and March 2022 as part of the MUSE programme 0108.B-0904 (PI: Fahrion) using the adaptive optics (AO)-supported wide-field mode \citep{AO, Galacsi}. As the AO system operates with strong sodium lasers, the wavelength range between 5800 and 5970 \AA\,around the Na D line is blocked. 
The observations were structured in observing blocks that implement a OSOOSO (O = object, S = sky) strategy, which brackets the on-target observations with offset sky pointings. Each science observation has 500 seconds of exposure time, while sky pointings have 150 seconds long exposures (see Table \ref{tab:galaxies}). Depending on the central surface brightness of our targets we acquired between one and four such observing blocks per target. For these galaxies, we reduced the raw data with the MUSE data reduction pipeline version 2.8.5 \citep{Weilbacher2020} within the \textsc{esorex} framework version 3.13.5. The reduction pipeline steps include bias correction, flatfielding, wavelength calibration, and sky subtraction.  

We complemented this sample with two archival galaxies NGC\,4592 and UGC\,8041 that were observed without AO. For those, we used the science-ready data available through ESO's science archive\footnote{\url{http://archive.eso.org/scienceportal/home}}. NGC\,4592 was originally observed as part of programme 095.B-0532 (PI: Carollo) in 2015 for the MUSE Atlas of Disks (MAD) survey \citep{ErrozFerrer2019}, while UGC\,8041 was observed as part of programme 0104.D-0503 (PI: Anderson) in 2020 as part of the ongoing supernova survey AMUSING\footnote{\url{https://amusing-muse.github.io/}}. 

In addition to being observed in our programme, NGC\,1487 was also observed as part of programme 0100.B-0116 (PI: Carollo) in 2018 with two MUSE pointings as part of the MAD survey with AO. To obtain a deeper data cube especially in the region around the NSC, we re-reduced this data and combined it with our 2021 observations, creating a three-pointing mosaic. 

\subsection{Galaxy sample}
The galaxies in our sample were selected among the limited sample of known nucleated dwarf LTGs observable with MUSE within reasonable time-constraints based on the catalogues reported by \cite{Boker2004}, \cite{Georgiev2009b}, and \cite{Georgiev2014}. All of these works used archival photometric data taken with the \textit{Hubble} Space Telescope (HST) in a diverse range of filters and instruments to study the nucleation fraction and properties of NSCs in local LTGs and irregular dwarf galaxies. The galaxies studied here are in isolated environments and have distances $\sim 5 - 20$ Mpc.
General properties of the galaxies and their NSCs are reported in Table \ref{tab:galaxies}, which lists distances, masses, foreground extinctions, and the spectrally averaged full width at half maximum (FWHM) of the point spread function (PSF) of each MUSE cube.  
Those were obtained by fitting a Moffat model to bright foreground stars at different wavelengths -- or the unresolved NSC if no stars were available -- in the field of view using \textsc{MPDAF} \citep{MPDAF}. The reported PSF FWHMs correspond to the average values between 4700 \AA\,and 7100 \AA, as the PSF varies with wavelength.

For visualisation of the galaxy sample, Fig. \ref{fig:galaxy_sample} presents RGB images obtained from synthetic $V$, $R$, and $I$-band images that were created with \textsc{MPDAF} out of the MUSE data for the nine galaxies in our sample. The inset in each panel shows a \mbox{10\arcsec $\times$ 10\arcsec\,}zoom around the NSC of each galaxy. We note that not all of the NSCs appear to lie directly in the photometric centre of these galaxies as some of them, such as ESO\,59-01 or UGC\,3755, are irregular dwarf galaxies with ill-defined centres (see also \citealt{Georgiev2009a, Georgiev2009b}), but the NSCs are the most massive star clusters in all of the galaxies and are located in the central regions, at least in projection. We further note that NGC\,1487 is an interacting system with an ongoing merger and the star cluster classified as the NSC appears to be part of a larger complex of star clusters and star-forming regions that appears to be displaced with respect to the main body of stars. As also the analysis of this star cluster shows (see following sections), its NSC is fundamentally different from those of the others in the sample. Nonetheless, we follow the classifications made in \cite{Georgiev2009b} and \cite{Georgiev2014} and analyse NGC\,1487 and its NSC alongside the rest of the galaxies, but we subsequently highlight this point where needed for the reader to keep in mind.

\section{Methods}
\label{sect:methods}
In the following, we describe how the spectra of NSCs, the circumnuclear regions, and the host galaxies were extracted from the data. Then we present how we obtained stellar ages, metallicities, SFHs and SFR maps.

\subsection{Extraction of spectra}
\label{sect:NSC_spectra}
Our main focus is to analyse the properties of NSCs in direct comparison to the central regions of their host galaxies and the hosts in general. To facilitate such a comparison, we consider the MUSE spectra of the NSC, the circumnuclear region and the host galaxies separately, as described in the following.

\subsubsection{Nuclear star cluster and circumnuclear spectra}

\begin{figure}
    \centering
    \includegraphics[width=0.48\textwidth]{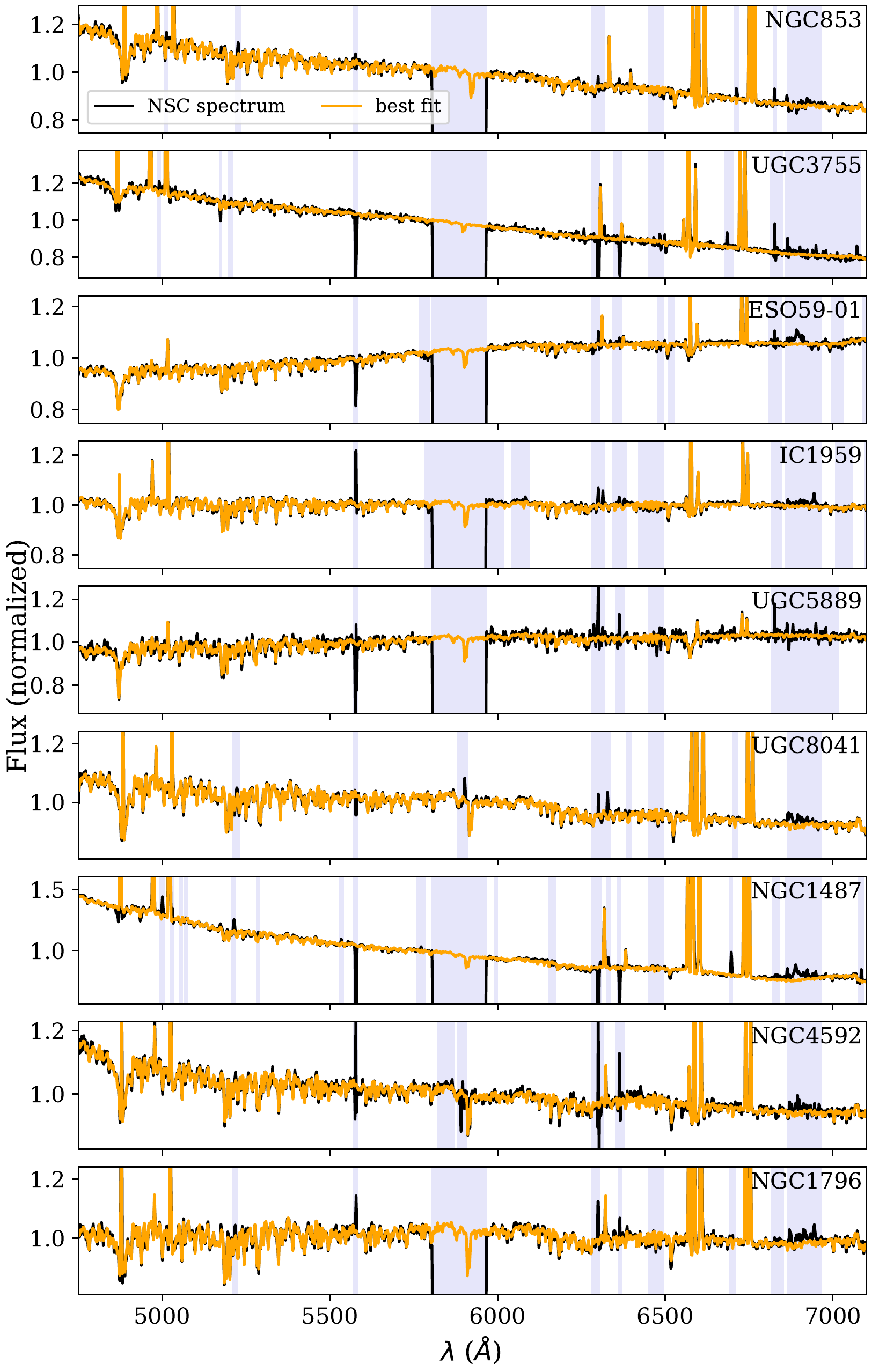}
    \caption{NSC spectra in black with \textsc{pPXF} fit in orange, showing a fit of both the absorption and emission lines. The light purple regions were masked during the fits. Each panel corresponds to one galaxy. We note that the y-axis scale is not the same in all panels. The range is chosen to highlight the absorption line features. For observations acquired with AO, the region around the Na D line is blocked.}
    \label{fig:NSC_spectra}
\end{figure}

We extracted the spectra of the NSCs using a PSF-weighted circular aperture centred on the NSC position in each MUSE cube. For this weighted extraction we used wavelength-dependent two-dimensional Moffat PSF models that were obtained with \textsc{MPDAF}. We note that all of these NSC spectra show the presence of emission lines with varying strength, as can be seen in the spectra shown in Fig. \ref{fig:NSC_spectra}. While the emission lines are very strong in galaxies such as NGC\,1487, NGC\,3755, or UGC\,853, other galaxies such as ESO\,59-01 and UGC\,5889 show only low level gas emission. We explore this further in Sect. \ref{sect:SFRs}.

For a direct comparison to the central galaxy regions, we further derived a background spectrum, using an annulus extraction with an inner radius of 8 (1.6\arcsec) and an outer radius of 13 pixel (2.6\arcsec), corresponding to an average distance of 2\arcsec\, from the NSCs. This extraction radius was chosen to be close to the NSC without being affected strongly by the light of the NSC to accurately determine the properties of the central regions of the galaxy. In the case of UGC\,8041 that has the broadest PSF FWHM, the inner annulus radius corresponds to $\sim3\sigma$ of the PSF. Choosing a slightly larger annulus radius does not affect the results. We refer to these spectra as the spectra of the circumnuclear region.

In \cite{Fahrion2021}, an additional step was to remove the galaxy contribution from the NSC spectrum by subtracting the background spectrum. In the present work, we decided against subtracting the background for several reasons. Firstly, due to the irregular nature of these dwarf galaxies, the background is complex and sometimes clumpy, especially in the gas structure. This makes an accurate accounting of the galaxy background challenging. Secondly, we note that NSCs in dwarf galaxies generally stand out significantly against the galaxy background, as can be seen from Fig. \ref{fig:galaxy_sample}. Finally, testing with background-subtracted NSC spectra revealed problems in the subsequent spectral fitting in UGC\,3755, which has the least massive NSC of our sample and we found that its background-subtracted NSC spectrum has an insufficient signal-to-noise (S/N) in the absorption lines for stellar population analysis.
Nonetheless, we verified for the other galaxies that the results remain unchanged even when the background spectrum is subtracted as also the main features in the SFHs presented in Sect. \ref{sect:results} are recovered. 

\subsubsection{Spectra of the host galaxy}
In addition to the NSCs and circumnuclear regions, we also consider the host galaxies directly. 
For a comparison of the SFHs, we binned the MUSE cubes using the \textsc{VorBin} method \citep{Cappellari2003} to a target S/N of 100. For NGC\,1487 we chose S/N = 200 due to the large size of the dataset. This binning ensures a S/N sufficient for an analysis of the galaxy SFHs in comparison to the NSC SFHs. In future work we will also use the binned cubes to analyse the galaxy stellar populations in more detail.

As described in Sect. \ref{sect:em_diagnostics}, we further used the original, unbinned MUSE cubes to obtain spaxel-by-spaxel maps of the SFR densities. Those are used to compare the star formation activity in the NSCs and their surroundings.

\subsection{Full spectrum fitting with pPXF}
In this work, we used \textsc{pPXF} \citep{Cappellari2004, Cappellari2017}, a full spectrum fitting routine that determines the line-of-sight velocity distribution as well as the distribution of ages and metallicities by fitting a spectrum with a library of stellar population synthesis models. We used the same setup of \textsc{pPXF} to fit the spectra of NSCs, circumnuclear regions, and the Voronoi-binned MUSE cubes. 

\subsubsection{Fit setup}
As in \cite{Fahrion2021}, we adopted the E-MILES single stellar population (SSP)\footnote{\url{http://miles.iac.es}} models \citep{Vazdekis2010, Vazdekis2016}. We used the models with a double power-law initital mass function with a high-mass slope of 1.30 \citep{Vazdekis1996} and BaSTi isochrones \citep{Pietrinferni2004, Pietrinferni2006} that span a regular grid of ages and total metallicities between 30 Myr and 14 Gyr and \mbox{[M/H] = $-2.27$ dex} and $+0.40$ dex, respectively. Due to this age grid, we are not able to constrain younger ages.
Given that the models are created from empirical spectra, they inherit the abundance pattern of their stellar library. Consequently, they have [Fe/H] = [M/H] at high metallicities and are $\alpha$-enhanced at lowest metallicities. The E-MILES models have a spectral resolution of 2.51\,\AA\,in the wavelength range covered by MUSE \citep{FalconBarroso2011}. To account for the instrumental resolution of MUSE that varies slightly with wavelength, we used the description of the MUSE line spread function from \cite{Guerou2016}.

We fitted the E-MILES models in a restricted wavelength range between 4700 and 7100 \AA\, to avoid contamination from residual sky lines at redder wavelengths. 
Any residual sky lines within this wavelength range were masked during the fitting. As recommended by \cite{Cappellari2004} and \cite{Cappellari2017}, we first performed initial fits to determine the line-of-sight velocity distribution using no multiplicative polynomials and additive polynomials of degree eight. Then, to determine the stellar population properties, we kept the kinematics fixed and fitted the spectra again using no additive polynomials, but multiplicative polynomials of degree 12. This ensures that the shape of absorption lines is preserved, but the multiplicative polynomials correct the shape of the continuum and thus account for reddening. When fitting for the stellar population properties, we masked the emission lines in the spectrum. 
However, as described in Sect. \ref{sect:SFRs}, we also performed fits in which absorption and emission lines were fitted simultaneously to obtain the line-of-sight velocities of the stellar and gaseous components in the NSCs (see Fig. \ref{fig:NSC_spectra}). Testing different fitting approaches showed that the stellar population results are mostly to whether the emission lines are masked or fitted simultaneously.

\subsubsection{Obtaining star formation histories}
As \textsc{pPXF} returns the weights of the models in the best fit, the mean age and metallicity can be constructed as weighted means, both mass- and light-weighted. Additionally, by applying regularisation that ensures a smooth distribution of weights, it is possible to derive SFHs as mass fractions versus age. In a regularised fit, \textsc{pPXF} returns the smoothest weight distribution that is still compatible with the data (see also \citealt{Cappellari2017}). As a consequence, not every small feature is reliable, but our testing has shown that the overall SFH shapes (single peak, multiple peaks, or extended) are robust against changes in the fit setup such as the considered wavelength range. Fitting in a broader wavelength range, for example up to 9000\,\AA\,recovers the same major features in the SFHs, but with larger uncertainties in the regions of strong sky residuals. However, we found that choosing a smaller wavelength range, for example 4700 -- 5600\,\AA\,leads to erroneous results for NGC\,1487 due to the very strong H$\beta$ emission line in this galaxy that is erasing much of the signal in the absorption line. In this galaxy, only a large wavelength range leads to a stable age solution, while in the small wavelength range, the grid boundaries were hit. For the other galaxies, the chosen wavelength range appears to have less of an effect on the overall SFH shape and the mean metallicity.
Also choosing the scaled-solar MILES models instead of the E-MILES models does not affect the overall results. However, with its lacking wavelength coverage bluer than 4700 \AA, the age accuracy of MUSE is limited, in particular for old stellar populations ($>$ 8 Gyr). 

\subsubsection{Obtaining mean values with statistical uncertainties}
We used a Monte-Carlo (MC) approach to obtain mean values of the kinematics and stellar population properties with their respective random uncertainties. Here, 100 representations of the spectrum were created by randomly perturbing the best-fit spectrum from an initial fit based on the respective residual (best-fit spectrum subtracted from the input spectrum). We used the resulting distribution of the fitted parameters to obtain mean ages and metallicities with their random uncertainties. Additionally, we also used this MC approach to obtain the line-of-sight velocity of the stellar and gaseous components in the NSC to search for any possible offsets. We note that this MC approach can only quantify statistical uncertainties, not systematic ones. 

\subsection{Emission line diagnostics}
\label{sect:em_diagnostics}
In addition to using the stellar absorption diagnostics, we used gas diagnostics to derive the star formation rates in the NSCs and their surroundings on a spaxel-by-spaxel basis.
To do so, we used the data analysis pipeline (\textsc{DAP})\footnote{\url{https://gitlab.com/francbelf/ifu-pipeline}} described in \cite{Emsellem2022}. This modular software framework creates  high-level data products out of integral-field spectroscopy data: maps of the stellar kinematics, stellar population properties, and emission line fluxes. For this paper, we focus on the SFR surface densities $\Sigma_{\text{SFR}}$ that are calculated from the H$\alpha$ fluxes. A more detailed analysis of emission line diagnostics will be presented in Bulichi et al., in prep. Furthermore, while \textsc{DAP} can also produce stellar population outputs, we fitted our spectra directly with \textsc{pPXF} as described above to ensure the same approach as in \cite{Fahrion2021}. We note that \textsc{DAP} is set up to work with a restricted SSP library covering 13 different ages and six different metallicities, but our tests show that the results from both approaches are in good agreement. In particular, kinematic parameters agree very well with each other, but due to the restricted library used in \textsc{DAP}, boundary effects can occur, leading to an artificial offset at low metallicities. 

As a first step in \textsc{DAP}, the MUSE data cubes are corrected for Milky Way foreground extinction using the values from Table \ref{tab:galaxies}. Then, the data are binned with the Voronoi binning method from \cite{Cappellari2003} to a target S/N of 90 and fitted for the stellar kinematics with \textsc{pPXF} to obtain  best-fitting spectra of the stellar continuum for each Voronoi bin. Emission line fluxes and velocities for each spaxel are then obtained using \textsc{pPXF} within \text{DAP}, which fits the emission lines with Gaussian templates after keeping the absorption line spectrum fixed with values obtained from the initial kinematic fits of the respective Voronoi bin of each spaxel. In this work, we only used H$\alpha$ and H$\beta$ to derive SFR surface densities and other results based on ionised gas diagnostics will be presented in future work. Also, a more detailed description of how SFRs are derived will be given in Bulichi et al., in prep., but we summarise here the key principles (see also \citealt{ErrozFerrer2019} or \citealt{Buzzo2021} for a description). 

First, we corrected the H$\alpha$ flux for internal dust extinction using the Balmer decrement (flux ratio H$\alpha$/H$\beta$), assuming an intrinsic ratio of 2.86 \citep{OsterbrockFerland2006} and the dust extinction law from \cite{Calzetti2000} for star-forming galaxies. Using the distances to each galaxy as in Table \ref{tab:galaxies}, extinction corrected fluxes were then converted to intrinsic H$\alpha$ luminosities $L_\text{int}(\text{H}\alpha)$. From those, the star formation rate in each spaxel was estimated using the relation from \cite{Hao2011}:
\begin{equation}
    \text{SFR}(M_\odot\,\text{yr}^{-1}) = 10^{-41.257} L_\text{int}(\text{H}\alpha),
\end{equation}
with $L_\text{int}(\text{H}\alpha$) in erg s$^{-1}$. This relation assumes a Kroupa IMF \citep{Kroupa2003} that is nearly identical to the double-power law IMF that we used for the stellar population analysis. By accounting for the pixel size of MUSE at the respective distance, we obtained maps of the star formation rate surface density in $M_\odot\,\text{yr}^{-1}$ kpc$^{-2}$. 


\begin{figure}
    \centering
    \includegraphics[width=0.45\textwidth]{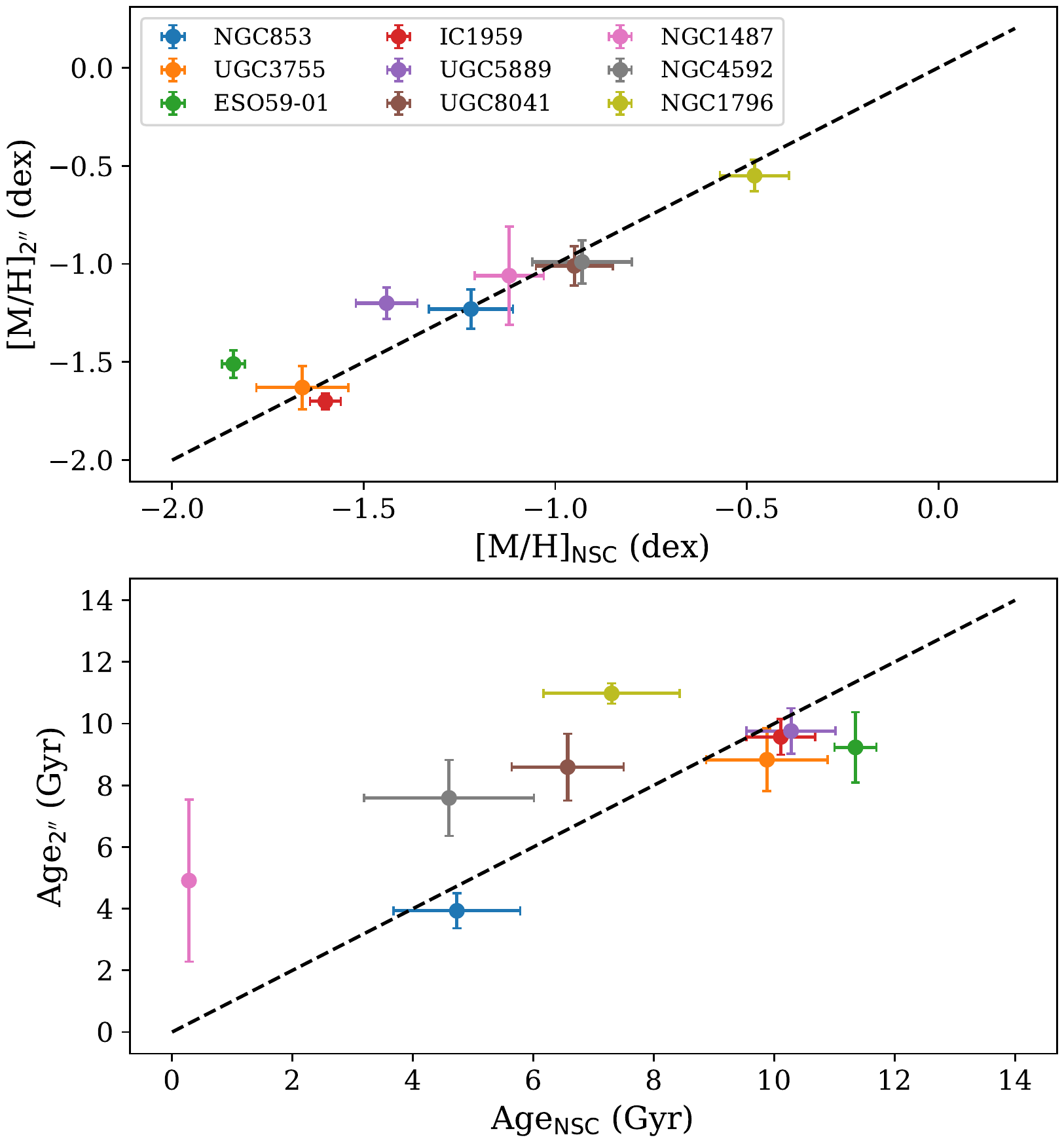}
    \caption{Comparison between metallicities (\textit{top}) and ages (\textit{bottom}) between NSC on x-axis and circumnuclear region on y-axis (see Table \ref{tab:fitting_results}). The dashed lines show the one-to-one relation.}
    \label{fig:age_metal_comp}
\end{figure}

\begin{figure*}
    \centering
    \includegraphics[width=0.92\textwidth]{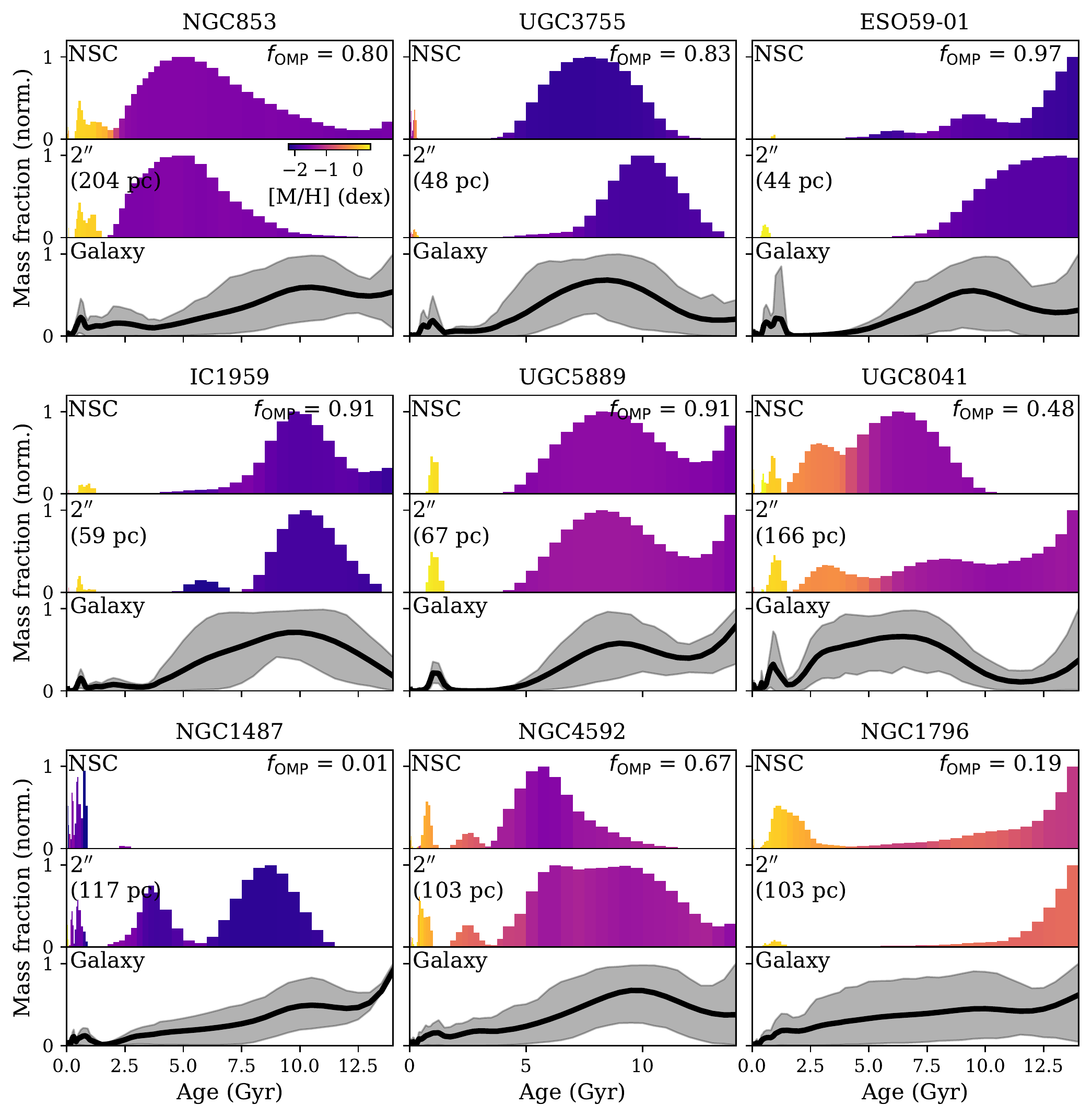}
    \caption{Star formation histories of the NSCs (\textit{top panel of the three panels for each galaxy}) from the NSC spectrum, at 2\arcsec\,radius from the NSC (\textit{middle}, the circumnuclear region), and the galaxies (\textit{bottom}, from Voronoi-binned spectra). The mass fractions were normalised such that the maximum is at 1 for visualisation. The NSC and 2\arcsec\, SFHs are colour-coded by the mean metallicity in each age bin to illustrate the age and metallicity distribution simultaneously. For the galaxies, we show the average SFH from the binned spectra as the thick black line and the 16th and 84th percentiles as grey shaded areas. The galaxies are ordered with increasing galaxy mass from the top left to the bottom right. The mass fraction of old, metal-poor populations ($> 2$ Gyr, $< -1.0$ dex) is indicated in the top right corner of each galaxy.}
    \label{fig:SFHs}
\end{figure*}

\begin{table*}
\caption{Nuclear star cluster fit results after MC fitting.}\vspace{-4mm}
\begin{center}
\begin{tabular}{c c c c c c} \hline\hline
Galaxy & $<$[M/H]$_{\text{NSC}}>$ & $<$Age$_{\text{NSC}}>$ & $<$[M/H]$_{\text{2\arcsec}}>$ & $<$Age$_{\text{2\arcsec}}>$ \\ 
 & (dex) & (Gyr) &  (dex) & (Gyr) \\ 
 (1) & (2) & (3) & (4) & (5)  \\ \hline
NGC\,853 & $-$1.22 $\pm$ 0.11 & 4.73 $\pm$ 1.05 & $-$1.23 $\pm$ 0.10 & 3.93 $\pm$ 0.57 \\
UGC\,3755 & $-$1.66 $\pm$ 0.12 & 9.88 $\pm$ 1.01 & $-$1.63 $\pm$ 0.11 & 8.83 $\pm$ 1.02 \\
ESO\,59-01 & $-$1.84 $\pm$ 0.03 & 11.35 $\pm$ 0.35 & $-$1.51 $\pm$ 0.07 & 9.23 $\pm$ 1.14 \\
IC\,1959 & $-$1.60 $\pm$ 0.04 & 10.11 $\pm$ 0.57 & $-$1.70 $\pm$ 0.04 & 9.57 $\pm$ 0.58\\
UGC\,5889 & $-$1.44 $\pm$ 0.08 & 10.28 $\pm$ 0.74 & $-$1.20 $\pm$ 0.08 & 9.76 $\pm$ 0.74 \\
UGC\,8041 & $-$0.95 $\pm$ 0.10 & 6.57 $\pm$ 0.93 & $-$1.01 $\pm$ 0.10 & 8.59 $\pm$ 1.08 \\
NGC\,1487 & $-$1.12 $\pm$ 0.09 & 0.28 $\pm$ 0.05 & $-$1.06 $\pm$ 0.25 & 4.91 $\pm$ 2.63 \\
NGC\,4592 & $-$0.93 $\pm$ 0.13 & 4.60 $\pm$ 1.41 & $-$0.99 $\pm$ 0.11 & 7.59 $\pm$ 1.23 \\
NGC\,1796 & $-$0.48 $\pm$ 0.09 & 7.30 $\pm$ 1.13 & $-$0.55 $\pm$ 0.08 & 10.98 $\pm$ 0.33 \\
\hline
\end{tabular}
\label{tab:fitting_results}
\end{center}
\tablefoot{(1) Galaxy name, (2), (3) average mass-weighted NSC metallicity and age, (4), (5) average mass-weighted metallicity and age of the host galaxy at 2\arcsec\,separation from the NSC. These results are plotted in Fig. \ref{fig:age_metal_comp}}
\end{table*}

\section{Results}
\label{sect:results}
We present our results in the following, firstly focussing on the average stellar population properties of the NSCs and their surroundings. Then we present the SFHs and discuss the origin of the emission lines.

\subsection{Ages and metallicities}
As we are interested in the mass assembly of the NSCs, we concentrate in the following on mass-weighted ages and metallicities. Table \ref{tab:fitting_results} lists the mean mass-weighted ages and metallicities of the NSCs and the circumnuclear region as obtained from fitting the NSC spectrum and the spectrum at 2\arcsec. The quoted uncertainties were obtained using the above described MC approach. Figure \ref{fig:age_metal_comp} shows the ages and metallicities of the NSCs and circumnuclear regions in a direct comparison.

Overall, the NSCs cover a broad range in metallicities and ages, similar to their host galaxies. In ESO\,59-01, NGC\,1487, UGC\,3755, and UGC\,5889 the NSC appears to be more metal-poor than the host galaxy, but due to the uncertainties only the difference in ESO\,59-01 is significant ($>$ 3$\sigma$, with $\sigma$ being the standard deviation). The NSCs of IC\,1959, NGC\,1796, NGC\,4592, and NGC\,8041 might be more metal-rich than the host galaxy at 2\arcsec, but again the difference is not significant. 

As Fig. \ref{fig:age_metal_comp} shows, the low-mass galaxies in our sample UGC\,3755, ESO\,59-01, IC\,1959, and UGC\,5889 tend to have metal-poor, old NSCs, with NGC\,853 being an exception of this. In contrast, the more metal-rich NSCs tend to be younger. In this figure, the presumed NSC in NGC\,1487 already reveals its peculiar stellar populations as the youngest star cluster in our sample with an age of only a few hundred Myr.

\subsection{Star formation histories}
To explore the NSC stellar population properties in more detail, we show the SFHs in Fig. \ref{fig:SFHs}. This figure compares the NSC SFH to the SFH at 2\arcsec radius and the average SFH of the respective host galaxy. For the latter, we combined the SFHs of all the binned spectra in each galaxy and show their average trend together with the 16th and 84th percentiles as shaded regions. The SFHs are presented as normalised mass fractions as a function of age and are consequently mass-weighted. We colour-code the NSC and circumnuclear (2\arcsec) region SFHs by the mean metallicity of each age-bin to illustrate their age and metallicity distribution. 

As in \cite{Fahrion2021} we find diversity in these SFHs. While several NSCs show a dominant peak at old ages ($> 2$ Gyr) and very low metallicities ($< -1.0$ dex, e.g. ESO59-01, IC\,1959, or UGC\,5889), the NSCs in NGC\,1796, NGC\,4592, or UGC\,8041 have SFHs with also higher metallicity and more prominent young populations. 

To quantify the relative importance of different populations, we list the mass fraction $f_\text{OMP}$ corresponding to old, metal-poor populations in the NSC (Age $>$ 2 Gyr, [M/H] $< -1.0$ dex) in each panel. Compared to the ancient GCs found in the Milky Way and other massive galaxies (e.g. \citealt{Leaman2013}), this age cut is placed at a relatively young age. This was chosen to also account for the finding that GCs in star-forming dwarf galaxies can generally be much younger than in massive galaxies (e.g. the Small Magellanic Cloud, \citealt{Bica2020}, or irregular dwarf galaxies \citealt{Puzia2008}). Additionally, this cut corresponds to the age limit down to which multiple populations in GCs in the Local Group are found, marking a distinction between GCs and young massive clusters \citep{Martocchia2018}. Similarly, the metallicity cut was chosen to reflect the typical low metallicities of GCs in dwarf galaxies (e.g. \citealt{Lamers2017}). For example, in the Large Magellanic Cloud (typical of the mass of our sample), these age and metallicity cuts would select the classical, massive GCs above the age gap (e.g. \citealt{DaCosta1991, Geisler1997}). We use this mass fraction as an indicator of how much mass might have been assembled through the inspiral of such GCs.

Comparing the NSC SFHs to the circumnuclear SFH at 2\arcsec\,separation shows that they agree in some cases, both in the distribution of ages and metallicities, for example in NGC\,853, ESO\,59-1, UGC\,5889 or IC\,1959. 
In the more massive galaxies in our sample, NGC\,1796, NGC\,4592, and UGC\,8041 the NSC SFH exhibits also younger populations than the surrounding galaxy. We note that the young populations found in the circumnuclear regions and the galaxies in general are also found in each of the NSC spectra, irrespective of the overall NSC populations. This highlights a tight connection between the NSC and the host galaxy, showing that star formation happening globally can affect the NSCs in these dwarf galaxies. NGC\,1487 stands out again with having only young ages in the SFH of the NSC.

\begin{figure}
    \centering
    \includegraphics[width=0.48\textwidth]{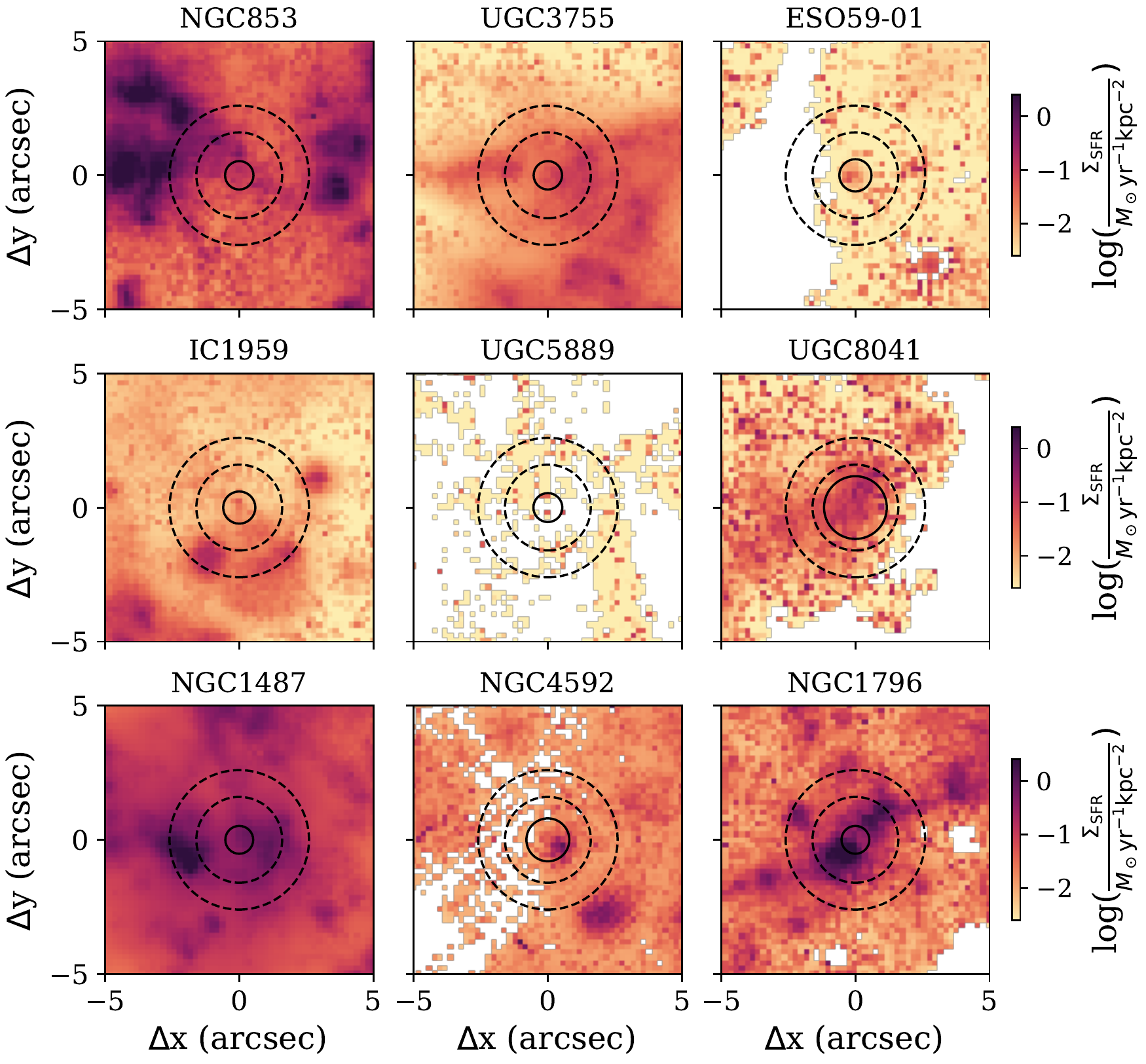}
    \caption{Star formation rate surface densities $\Sigma_{\text{SFR}}$ for the different galaxies. Each panel shows the 10\arcsec $\times$ 10\arcsec\, zoom to the region around the NSCs. The positions of the NSCs are highlighted by the circles. Each circle has a radius corresponding to the PSF FWHM. The dashed circles highlight the circumnuclear region at 2\arcsec, defined by an annulus with inner radius of 8 pixel = 1.6$\arcsec$ and outer radius of 13 pixel = 2.6$\arcsec$. Regions with low signal-to-noise in the H$\alpha$ flux were masked.}
    \label{fig:SFRs}
\end{figure}

\begin{table*}
    \caption{Comparison between gas and stellar velocity.}\vspace{-4mm}
    \begin{center}
    \begin{tabular}{c c c c c c c c}\hline\hline
         Galaxy & v$_{\text{NSC, stars}}$ & v$_{\text{NSC, gas}}$ & v$_{\text{2\arcsec, stars}}$ & v$_{\text{2\arcsec, gas}}$  & <log$(\Sigma_{\text{SFR, NSC}})$> & <log$(\Sigma_{\text{SFR, 2\arcsec}})$>  \\
          & (km s$^{-1}$) & (km s$^{-1}$) & (km s$^{-1}$) & (km s$^{-1}$) & log($M_\sun$ yr$^{-1}$ kpc$^{-2}$) & log($M_\sun$ yr$^{-1}$ kpc$^{-2}$) \\ 
          (1) & (2) & (3) & (4) & (5) & (6) & (7) \\
          \hline
        NGC853 & 1534.2 $\pm$ 2.0 & 1552.9 $\pm$ 0.7 & 1528.3 $\pm$ 2.2 & 1542.7 $\pm$ 0.4 & $-$0.94 $\pm$ 0.12 & $-$1.01 $\pm$ 0.40\\
        UGC3755 & 316.2 $\pm$ 4.8 & 302.9 $\pm$ 0.4 & 316.7 $\pm$ 4.1 & 305.0 $\pm$ 0.5 & $-$1.31 $\pm$ 0.17 & $-$1.57 $\pm$ 0.31\\
        ESO59-01 & 527.2 $\pm$ 1.3 & 524.0 $\pm$ 0.5 & 529.4 $\pm$ 3.5 & 531.9 $\pm$ 0.7 & $-$2.09 $\pm$ 0.35 & $-$2.52 $\pm$ 0.30 \\
        IC1959 & 645.4 $\pm$ 1.3 & 648.2 $\pm$ 0.5 & 646.0 $\pm$ 2.3 & 647.2 $\pm$ 0.4 & $-$1.92 $\pm$ 0.21 & $-$1.87 $\pm$ 0.46\\
        UGC5889 & 568.4 $\pm$ 2.0 & 577.7 $\pm$ 2.4 & 576.2 $\pm$ 3.1 & 568.1 $\pm$ 4.4 & $-$2.58 $\pm$ 0.85 & $-$3.17 $\pm$ 0.10\\
        UGC8041 & 1320.6 $\pm$ 1.3 & 1323.2 $\pm$ 0.9 & 1322.1 $\pm$ 1.9 & 1328.1 $\pm$ 1.2 & $-$1.15 $\pm$ 0.30 & $-$1.79 $\pm$ 0.50\\ 
        NGC1487 &  838.0 $\pm$ 2.5 & 851.6 $\pm$ 2.0 & 840.6 $\pm$ 3.4 & 841.6 $\pm$ 1.4 & $-$0.34 $\pm$ 0.14 & $-$0.62 $\pm$ 0.23\\
        NGC4592 & 1061.9 $\pm$ 1.6 & 1061.3 $\pm$ 0.6 & 1059.2 $\pm$ 1.3 & 1061.9 $\pm$ 0.6 & $-$1.38 $\pm$ 0.52 & $-$1.82 $\pm$ 0.19\\
        NGC1796 & 1015.8 $\pm$ 1.4 & 1019.2 $\pm$ 0.3 & 1017.8 $\pm$ 1.0 & 1012.3 $\pm$ 0.4 & $-$0.12 $\pm$ 0.43 & $-$1.10 $\pm$ 0.41\\
        \hline
    \end{tabular}
\end{center}
\tablefoot{(1) Galaxy name, (2), (3) average line-of-sight velocity of the stellar and gaseous component in the NSC spectrum, (4), (5) average line-of-sight velocity of stellar and gaseous components in the circumnuclear spectrum (2\arcsec), (6), (7) average SFR density in the NSC (defined by a circular aperture with radius corresponding to the PSF FWHM) and at 2\arcsec (defined by an annular aperture).}
\label{tab:gas_results}
\end{table*}

\subsection{Origin of emission lines}
\label{sect:SFRs}

As seen in Fig. \ref{fig:NSC_spectra}, all NSC spectra display conspicuous emission lines that are indicative of ongoing star formation. To explore the origin of these emission lines and investigate correlation between ionised gas distributions in the vicinity of the NSCs, we show maps of SFR surface density in a 10\arcsec $\times$ 10\arcsec\,cut-out around each NSC in Fig. \ref{fig:SFRs}. The overall distribution of SFR density in each galaxy will be discussed in future work. Here we focus on the nuclear regions and use the same colour scaling for all galaxies to emphasise the differences in the SFR densities between the galaxies. While galaxies such as ESO\,59-01 or UGC\,5889 generally show low levels, NGC\,853 and NGC\,1487 reach SFR densities of up to $\sim 0.4 \, M_\sun \text{yr}^{-1} \text{kpc}^{-2}$. 

Interestingly, many of the NSCs do not appear as distinguishable features in the SFR density maps, showing that the NSCs are not the regions with the most intense star formation in these galaxies. The NSCs in IC\,1959, UGC\,3755, and NGC\,853 lie in or at the borders of regions with star formation, but are not visible as separate peaks, whereas they clearly stand out in the colour images of Fig. \ref{fig:galaxy_sample}. Similarly, the NSC in NGC\,4592 is located right next to a peak in the SFR density map and the NSCs of NGC\,1796 and NGC\,1487 appear to lie between brighter knots of star formation. Only in ESO59-01 the NSC appears as a compact peak in the SFR density map, and the NSC in UGC\,8041 also appears to be star-forming, although the lower spatial resolution of this data makes it difficult to interpret the origin of the central emission. Nonetheless, the presence of these peaks is somewhat unexpected as the SFH of the NSC in ESO\,59-01 indicates mainly an old, metal-poor population and similarly, the NSC of UGC\,8041 is overall old and metal-rich. However, at closer inspection both galaxies host a small fraction of young stars that could be connected to the observed peaks in the SFR density. While a comparison between stellar and gas-phase metallicity is not trivial, we note here that the nuclear region in ESO59-01 exhibits low gas-phase metallicities and high gas-phase metallicities in UGC\,8041. A more detailed analysis will be presented in Bulichi et al., in prep.

To quantify the level of star formation in the NSC and its immediate surroundings, Table \ref{tab:gas_results} reports the average SFR densities in a circular aperture around the NSC as well as those measured within an annular aperture at 2\arcsec. We used the PSF FWHM of each MUSE cube as the radius of the circular aperture, corresponding to the light of an unresolved point-source such as the NSCs. As can be seen, in some galaxies, the average SFR density is higher in this central aperture than in the surroundings, for example in NGC\,1796, NGC\,1487, and NGC\,8041, while in others it is on a similar level. However, the maps show how complex the distribution in the SFR density is, so the comparison of these averaged values is not straight-forward.

To explore whether the emission lines present in all of the NSC spectra are truly originating from star formation within the NSC as opposed to just along the line-of-sight, we can further investigate whether the stellar and gas velocities show any significant offset that could indicate different physical origins. We report these measurements based on 100 MC fits to spectra of NSCs and the circumnuclear regions in Table \ref{tab:gas_results}.
For these fits, we fitted for the stellar and gaseous velocities simultaneously.

A comparison of the stellar velocities of the NSCs and the circumnuclear regions reveals no significant offsets, indicating that the NSCs are kinematically at rest with respect to the central regions of the galaxies, giving further indications that these star clusters truly are NSCs. However, NGC\,1487 and NGC\,853 show significant offsets ($> 3 \sigma$) between the stellar and gas velocity in the NSC, indicating a different physical origin of these two components. Additionally, the stellar and gaseous velocities obtained from the circumnuclear spectrum at 2\arcsec\, show a significant offset in NGC\,1796 and NGC\,853. As this spectrum is obtained by averaging over a larger area, the interpretation of this offset is a bit more challenging, but could indicate a turbulent gas velocity field. Due to the limited spectral resolution of MUSE, we do not explore the velocity dispersion as they are close or below the instrumental resolution of MUSE. Additionally, the spectral resolution is insufficient to determine whether the emission lines contain multiple components that would, for example, indicate outflows.

\section{Discussion}
\label{sect:discussion}
In the following, we discuss our results for each galaxy individually to infer the dominant NSC formation channel. Then, we compare our findings to recent results from the literature.

\subsection{Inferring the dominant NSC formation channel}
\label{sect:dominant_channel}
We aim to infer the dominant NSC formation channel for each of the nine galaxies studied here, differentiating between formation through accretion and mergers of GCs and central star formation. As described in the introduction, NSCs formed purely from GCs can show low metallicities and simple SFHs with old ages, whereas NSC formation through star formation occurring in situ at the galaxy centre directly can lead to more complex SFHs and the presence of young stars or ionised gas if the star formation is still ongoing. 

To distinguish between the two channels, we consider the mean ages, metallicities, and SFHs of the NSCs in comparison to the circumnuclear regions and the galaxies in general. We consider populations with Age $>$ 2 Gyr and [M/H] $<$ $-$1.0 dex as possibly stemming from GCs, quantified in the mass fraction of such populations contributing to the NSC mass $f_\text{OMP}$. The age cut was chosen, as described above, to reflect the generally broad age range of GCs in star-forming dwarf galaxies (e.g. \citealt{Puzia2008}). Additionally, numerical simulations and semi-analytical models have shown that dynamical friction leads to orbital decay of massive ($\sim 10^5 - 10^6 M_\sun$) star clusters on timescales of a few Gyr (e.g. \citealt{CapuzzoDolcetta1993,  CapuzzoDolcetta2008, CapuzzoDolcetta2008a, Agarwal2011}), making it possible to find such populations in a NSC formed from GCs.
The metallicity cut was made at such low metallicities as GCs in dwarf galaxies are typically characterised by a metallicity well below the average metallicity of their host \citep{Lamers2017}, also evident from their blue colour distributions (e.g \citealt{Peng2006}). We assume here that accreted GCs have the same stellar population properties as GCs still outside the centre, although more massive GCs are more likely to inspiral due to shorter dynamical friction timescales (e.g. \citealt{Tremaine1975}). However, given that GCs only show a very weak relation between mass and metallicity (e.g. \citealt{Fensch2014, Zhang2018, Fahrion2020c}), accreted GCs should have the same stellar population properties as their lower-mass counterparts.

For populations with higher ages and metallicities, we assume that they were likely formed more recently, from pre-enriched material close to the galaxy centre. Due to the unresolved nature of our study, we cannot derive whether this star formation has occurred directly in the NSC, which would be true in situ formation, or whether these populations stem from young clusters that formed in the vicinity of the nucleus and quickly spiralled inwards due to their short dynamical friction timescales (e.g. \citealt{Nguyen2014, Guillard2016}). As also described in the introduction, we understand both of these possibilities as formation from central star formation that clearly distinguishes them from accretion of old, metal-poor GCs.

We find that this fraction of mass in old, metal-poor components ($f_{\text{OMP}}$) is tightly correlated with the relative age difference between the NSC and circumnuclear region ($\Delta_{\text{age}} \equiv \text{Age}_\text{NSC}-\text{Age}_{2\arcsec}$). As an instructive example, the NSCs with the highest $f_\text{OMP}$ values show ages which are 1 -- 3 Gyrs older than the surrounding stars, while the very lowest $f_{\text{OMP}}$ NSCs have ages younger than the circumnuclear region by 2 -- 4 Gyrs. The correlation between $f_{\text{OMP}}$ and $\Delta_\text{age}$ crosses a value of $f_\text{OMP}$ = 0.5 where $\Delta_\text{age} \sim 0$, showing that a composite formation pathway implied by $f_\text{OMP}$ is consistent with a negligible age difference between the NSC and its surroundings. This shows that $f_\text{OMP}$ is a reasonable choice to quantify the dominant NSC formation channel, but to further refine the age and metallicity cuts, we aim to further analyse the star clusters of these galaxies in future work.

\subsubsection{NGC853}
According to mass measurements from the literature, NGC\,853 is the least massive galaxy of our sample, but has a massive NSC for its mass that constitutes $\sim 22 \%$ of the total stellar mass (see Table \ref{tab:galaxies}). A comparison to NSC-to-galaxy mass ratios from \cite{Neumayer2020} shows that this value is not particularly extreme as very low-mass galaxies ($M_\ast \sim 10^6 M_\sun$) can reach even $>$ 30\%. However, we note that these mass ratios were obtained for early-type dwarf galaxies.

The NSC of NGC\,853 is located adjacent to a region with strong star formation, but the gas and stellar velocities show a significant offset of $\Delta v = 8.8 \pm  2.1$ km s$^{-1}$, suggesting that the emission lines in its spectrum are not originating from currently ongoing star formation in the NSC. 
The SFH of NGC\,853's NSC is complex and extended, with a dominant peak at ages $\sim$ 5 Gyr and low metallicities, but it is not as metal-poor as the NSCs of ESO\,59-01 or IC\,1959, for example. The old and metal-poor components ($> 2$ Gyr, $< -1.0$ dex) make up 80\% of the NSC mass. Additionally, the NSC shows the presence of a young, metal-rich population. Given the dominant old and metal-poor population, we conclude that this NSC has formed most of its mass from accretion of GCs, however recent enriched star formation has also further shaped this NSC.

\subsubsection{UGC3755}
The NSC of UGC\,3755 is located in a region that shows star formation, but the NSC does not appear as a separate peak in Fig. \ref{fig:SFRs}. In addition, there is a velocity offset between the stellar and gaseous components in the NSC of $\Delta v = 13.3 \pm 4.8$ km s$^{-1}$. This velocity offset is not significant and a higher S/N spectra would be needed to confirm it. No offset is found between the stellar velocity of the NSC and the circumnuclear region, suggesting that the NSC is at rest with respect to the stellar component of UGC\,3755.

The SFH of UGC\,3755's NSC shows a dominant, old and metal-poor population in addition to a minor young and enriched population. With a mean age of $\sim 10 \pm 1$ Gyr and mean metallicity of $-$1.60 $\pm$ 0.13 dex, this NSC has a similar low metallicity as other star clusters in UGC\,3755 analysed using FORS2 spectra, but appears to be older as at least some of the GCs (2 - 6 Gyr) in this galaxy \citep{Puzia2008}. Based on the low metallicity and the comparison to other GCs in UGC\,3755, we conclude that the NSC in UGC\,3755 has assembled most of its mass through GCs.

\subsubsection{ESO59-01}
ESO\,59-01 shows low levels of star formation across the galaxy (see Bulichi in prep.), but we find a small peak of star formation at the NSC position. The gaseous and stellar components in the NSC agree in their line-of-sight velocities, further suggesting that this ionised gas is associated with low-level of star formation currently occurring in the NSC.

The NSC SFH is clearly dominated by an old and metal-poor component ($> 2$ Gyr, $< -1.0$ dex) that comprises 97\% of the mass and the NSC is more metal-poor than the circumnuclear region. Consequently, it is likely that the NSC in ESO\,59-01 has assembled most of its mass early-on through the accretion and mergers of GCs. However, the ongoing star formation and the minor young population in the NSC suggest that both channels have contributed to its formation, although with strongly unequal shares. Consequently, we identify the GC-accretion channel to be the dominant one in the formation of this NSC.

\subsubsection{IC1959}
Compared to ESO\,59-01, IC\,1959 shows much more star formation across its central region (Fig. \ref{fig:SFRs}). The NSC is not seen as a peak in star formation, but is located next to star-forming clumps. The velocities of the gaseous and stellar component in its spectra agree, suggesting that the ionised emission lines in the NSC spectrum are originating from low-level star formation in the NSC.

Similar to ESO\,59-01, the SFH is strongly dominated by an old and metal-poor component constituting 91\% of the NSC mass. Additionally, a minor young and metal-rich component is found. Consequently, also this NSC has likely formed most of its mass through GCs, but the presence of emission lines connected spatially and velocity-space to the NSC indicate additional contributions from low-level central star formation.

\subsubsection{UGC5889}
Out of the galaxies in our sample, UGC\,5889 shows the least amount of star formation across the MUSE field of view and its NSC is located in a region with a very low SFR density. Furthermore, the NSC spectrum shows weak emission lines and the kinematic analysis revealed an non-significant offset of $\Delta v = 9.1 \pm 3.1$ km s$^{-1}$. 

The NSC SFH shows predominantly old ages at low metallicity. This component constitutes 91\% of the mass, but also a minor young and metal-rich component is present. Hence, we conclude that this NSC was formed predominantly through the GC accretion channel.

\subsubsection{UGC8041}
UGC\,8041 shows knots of star formation along its spiral arms as well as at the NSC position. Since the gas and stellar velocities agree, it is likely that the star formation originates from within the NSC. 

The NSC SFH exhibits two populations, an older, metal-poor comprising 48\% of the mass and a young and enriched population. Based on the ongoing star formation and the presence of both old and young stars, we conclude that both channels have contributed significantly to the formation of this NSC.

\begin{figure*}
    \centering
    \includegraphics[width=0.95\textwidth]{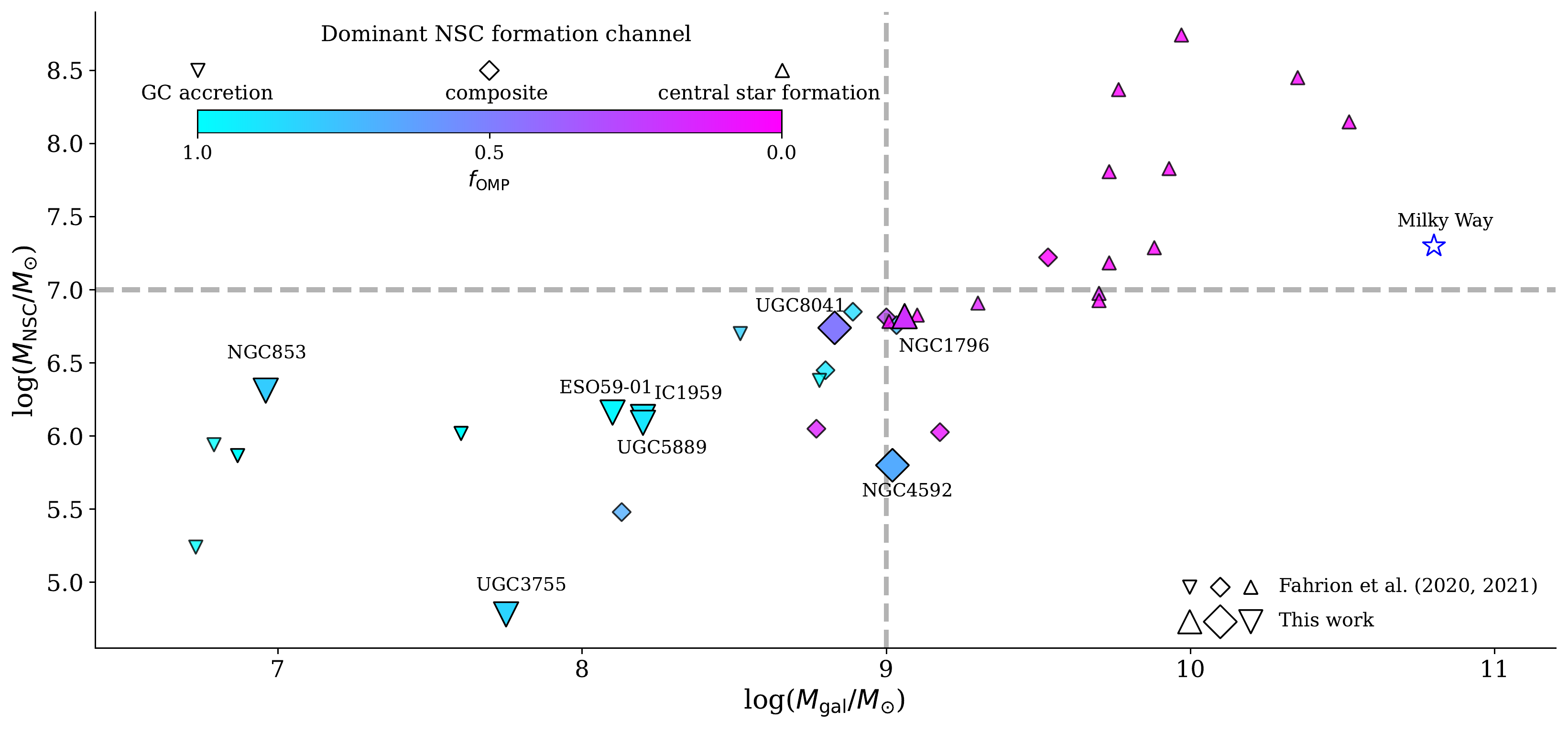}
    \caption{Dominant NSC formation channel for the studied galaxies on the NSC mass-galaxy mass plane. The large different symbols indicate the identified dominant NSC formation channel for the galaxies analysed in this work. The smaller symbols refer to the results presented in \cite{Fahrion2020, Fahrion2021}. We colour-coded the symbols by the mass fraction of old and metal-poor populations ($> 2$ Gyr, $< -1.0$ dex, see Table \ref{tab:fitting_results}). The star shows the position of the MW NSC for reference \citep{Neumayer2020}. The dashed lines indicate the mass limits where a transition of the dominant NSC formation channel was proposed. NGC\,1487 was excluded from this figure.}
    \label{fig:dominant_channel}
\end{figure*}

\subsubsection{NGC1487}
NGC\,1487 is currently undergoing a merger, leading to high levels of star formation and a strongly distorted morphology (e.g. \citealt{Buzzo2021}). The star cluster classified as its NSC is visually removed from the centre of the old stellar populations in either of the merger remnants. This star cluster is located in one of the regions that show the highest levels of star formation, but there is no clear peak in the SFR density map at its location. Additionally, there seems to be a significant velocity offset between the gaseous and stellar component in the NSC spectra of $\Delta v = 13.6 \pm 3.3$ km s$^{-1}$. This indicates that the emission lines in the NSC spectra are not arising from the NSC itself.

The SFH of NGC\,1487's NSC is the only case where we do not find any indications of a population older than 3 Gyr. Instead, the SFH only shows ages below 3 Gyr with an mean age at $\sim$ 300 Myr, much younger than the background.
However, the metallicity of these newly formed star clusters is very low for its age ($-$1.12 dex).  This is typical of triggered star clusters during mergers which bring in low metallicity gas from the outskirts, as typical young stars forming in the central regions of dwarf galaxies of this mass have higher metallicities.

Interestingly, the zoomed inset in Fig. \ref{fig:galaxy_sample} shows that this star cluster is surrounded by several bright sources. Comparing with the SFR density map in Fig. \ref{fig:SFRs} reveals that one of these sources is likely a star-forming region, but the others appear to have no counterpart in the SFR map. An analysis of the other star clusters next to the central one reveals similarly young ages of a few hundred Myr and low metallicities, possibly indicating that the complex of star clusters formed during the merger. Based on the anomalous chemistry, young age, surrounding clusters and distance from the optical centre of the old stars it is likely the star cluster classified as the NSC of NGC\,1487 is a fundamentally different object than the centrally located NSCs in our sample, whose high densities reflect formation processes which operate specifically due to the NSC living at deepest point of the gravitational potential well. Instead of being formed through one of the discussed NSC formation channels, we propose that this star cluster complex in NGC\,1487 has formed recently, likely in the ongoing merger. 
\subsubsection{NGC4592}

NGC\,4592 shows strong star formation across the galaxy's disk. The NSC does not appear as a peak in the maps of SFR density, but is located directly next to one. The velocities of gaseous and stellar components in the NSC agree with each other, making it likely that the ionised gas is originating from star formation within the NSC.

We note here that even though NGC\,4592 has a very similar mass to NGC\,1796, its NSC is less massive by roughly one order of magnitude. The galaxy and NSC metallicity in NGC\,4592 is lower than in NGC\,1796, but its NSC has a higher metallicity than many of the others in the sample. Its SFH is dominated by young and intermediate ages ($\sim$ 5 Gyr). The fraction of stars with ages $>$2 Gyr and a metallicity $< -1.0$ dex constitute 67\% of the NSC mass. For NGC\,4592, we conclude that both formation channels have contributed significantly to the formation of this NSC.

\subsubsection{NGC1796}
NGC\,1796 is the most massive galaxy in our sample and shows high levels of star formation throughout its disk and bar region. The NSC is located in the bar and while it does not appear as a separable peak of star formation, the velocities of the gaseous and stellar components in the NSC agree, indicating that the star formation occurs within the NSC. 

The NSC in NGC\,1796 is the most metal-rich of the sample and it is dominated by young ages. The fraction of old, metal-poor stars is 19\%. For these reasons, we conclude that the NGC\,1796's NSC has build most of its mass through central star formation. Additional accretion from GCs appears to be minor, but cannot be excluded.

\subsection{NSC formation as function of galaxy mass}
Out of the nine galaxies in our limited sample, five appear to have formed most of their NSC mass through the mergers of old, metal-poor GCs. This formation channel is suggested by the large value of $f_\text{OMP}$ in the recovered mass fractions, and the positive $\Delta_\text{age}$ values suggesting more metal-poor GCs are the dominant contributor to the composition of these NSCs. In contrast, the NSC of NGC\,1796 was likely formed predominantly through central star formation as evident from the younger populations and a low value of $f_\text{OMP}$. Likely both channels contributed to the formation in UGC\,8041 and NGC\,4592 where the SFH shows both a old, metal-poor and significant young, metal-rich component, with intermediate mass fractions in old metal poor stars, and $\Delta_\text{age}$ close to zero. In the merging galaxy NGC\,1487, the complex of newly formed massive star clusters have low metallicity and likely were triggered by infalling gas in the merger process.  These different objects are also anomalous in that they show no old populations -- consistent with massive star clusters formed recently in the field, rather than a classical NSC residing at the barycentre of a system where continued star formation or cluster accretion can proceed.

To explore these results in the context of a larger sample of galaxies, we show them on the plane of galaxy and NSC mass in Fig. \ref{fig:dominant_channel}. In this figure, we also included the 25 early-type galaxies studied with similar methods by \cite{Fahrion2021} as well as the two early-type dwarfs KK\,197 and KKs\,58 that were observed with MUSE and analysed in a similar way in \cite{Fahrion2020}. This figure shows the inferred dominant NSC formation channel with different symbols and colours. We applied the same cut in age ($>$ 2 Gyr) and metallicity ($< -1.0$ dex) to these results to colour-code the symbols by $f_\text{OMP}$. Due to its different nature, we have not included NGC\,1487.

Our present findings agree well with the results from \cite{Fahrion2020, Fahrion2021} and support a trend in the dominant NSC formation channel with galaxy mass. NGC\,1796, in which we found evidence of the central star formation scenario, is the most massive in our sample with $M_\text{gal} \sim 10^9 M_\sun$, where \cite{Fahrion2021} suggested that the transition between GC accretion and central star formation occurs (see also \citealt{Neumayer2020, Pinna2021}). UGC\,8041 and NGC\,4952 that have both similarly high masses appear to have formed their NSCs through both channels.
In the lower mass galaxies, we identified the GC accretion channel to be responsible for most of the mass in their NSCs in five galaxies, similar to the findings of \cite{Fahrion2020, Fahrion2021}, as indicated by higher values of $f_\text{OMP}$. This is an interesting result as the galaxies studied here show star formation globally while the sample of \cite{Fahrion2020} and \cite{Fahrion2021} only considered early-type dwarfs even though some of them showed the presence of ionised gas confined to the central regions.

Studies of massive ($M_\ast > 10^9 M_\sun$) star-forming galaxies have found that their NSCs are mostly formed through central star formation; although, signatures of old and metal-poor components and hence GC accretion have also been found \citep{Kacharov2018, Hannah2021}. The NSC of the Milky Way is here a prominent example as it shows both young stars likely formed in situ and old, metal-poor components that possibly were accreted from GCs \citep{Antonini2012, FeldmeierKrause2015, FeldmeierKrause2020, ArcaSedda2020}. Similarly, the massive ETGs studied in \cite{Fahrion2021} show high metallicities and sometimes young ages in their NSCs, indicating that their formation was dominated by central star formation.

Consequently, the results in this work support a consistent picture in which low-mass galaxies ($M_\text{gal} < 10^9 M_\sun$) form their NSCs predominantly through GC accretion, while the mass growth of NSCs in more massive galaxies appears to be dominated by enriched star formation occurring in the central regions directly -- irrespective of whether the galaxies are star-forming or not. This finding is also in agreement with the semi-analytical model presented in \cite{Fahrion2022} that considers NSC formation through GC accretion and finds that the massive NSCs found in massive galaxies require significant growth through central star formation to explain their high masses. Nonetheless, accretion of GCs might also seed the NSCs of these massive galaxies that then later grow their NSCs further through central star formation.

We note here that even though we have identified GC accretion to be the dominant NSC formation channel in the low-mass galaxies, their NSCs still show some minor young and metal-rich components. This illustrates the complex interplay between NSC formation and galaxy evolution: even though the NSCs in dwarf galaxies of different morphological types form predominantly through inspiralling GCs, the detailed properties and SFHs of the NSCs depend on their host galaxy's evolution. We note that these minor populations were also found when the background-subtracted NSC spectrum was used, indicating that they are truly originating from within the NSC as opposed to be caused by a contamination of the spectrum from the galaxy contribution.

\section{Conclusions}
\label{sect:conclusions}
This paper presents the first analysis of nucleated late-type dwarf galaxies with integral-field spectroscopy. Using MUSE data of nine galaxies, we find:
\begin{itemize}
    \item All galaxies present some level of star formation with its intensity varying across each galaxy and between galaxies in the sample. 
    \item Emission lines from ionised gas are observed in the integrated light of all NSCs.
    \item To explore the origin of the emission lines and establish if they are related to NSCs, we analyse maps of SFR density surrounding the NSCs. Most NSCs do not have a distinct peak in these maps. Furthermore, we find a significant offset in line-of-sight velocities of the gaseous and stellar components in the NSC spectrum in two of the galaxies. This suggests that the emission lines in the NSC spectra are not always associated with the NSCs themselves, but can originate from gas located along the line-of-sight towards the NSC.
    \item The NSCs span a broad range in metallicities and ages. 
    NSCs of low-mass galaxies are more metal-poor and older than their surrounding areas - hinting at the role of GC in-spiral. NSCs in progressively higher mass galaxies have average ages and metallicities which are younger and more enriched than the circumnuclear region.
    \item The NSC SFHs are diverse in their shape. Five of the nine galaxies have SFHs dominated by old and metal-poor populations, indicative of formation through accretion of metal-poor GCs. Additionally, the SFHs and metallicity distributions in three of the most massive galaxies in our sample are extended or metal-rich, indicating NSC formation through central-star formation. 
    \item Based on the SFHs, metallicity distributions and the presence of star formation, we identified the most likely dominant NSC formation channel in the nine galaxies. In the most massive galaxy of our sample, we concluded that its NSC was formed predominantly through central star formation, whereas we identified the GC accretion channel as the dominant NSC formation pathway in five galaxies. In two galaxies, both channels appear to be contributing.
    \item Our results support suggested scenarios from previous studies using similar methods, in which GC accretion builds most of the mass in NSCs of low-mass galaxies whereas central star formation becomes more relevant for more massive galaxies $M_\text{gal} \geq 10^9 M_\sun$. 
\end{itemize}
Our analysis supports a scenario where NSCs in dwarf galaxies form most of their mass through the accretion of GCs, even though star formation at a low level can further shape the NSCs in late-type dwarfs. Given our limited sample of nine galaxies, future investigation of larger samples will be crucial to establish how galaxy morphology and evolution can affect the formation of NSCs.

\begin{acknowledgements}
We thank the anonymous referee for helpful comments. KF acknowledges support through the ESA research fellowship programme. TEB acknowledges support through the Leiden/ESA Astrophysics Program for Summer Students (LEAPS) 2022. O.M. is grateful to the Swiss National Science Foundation for financial support under the grant number PZ00P2\_202104. GvdV acknowledges funding from the European Research Council (ERC) under the European Union's Horizon 2020 research and innovation programme under grant agreement No 724857 (Consolidator Grant ArcheoDyn).

\end{acknowledgements}

\bibliographystyle{aa} 
\bibliography{References}

\begin{thebibliography}{79}
\expandafter\ifx\csname natexlab\endcsname\relax\def\natexlab#1{#1}\fi

\bibitem[{{Agarwal} \& {Milosavljevi{\'c}}(2011)}]{Agarwal2011}
{Agarwal}, M. \& {Milosavljevi{\'c}}, M. 2011, \apj, 729, 35

\bibitem[{{Antonini}(2013)}]{Antonini2013}
{Antonini}, F. 2013, \apj, 763, 62

\bibitem[{{Antonini} {et~al.}(2015){Antonini}, {Barausse}, \&
  {Silk}}]{Antonini2015}
{Antonini}, F., {Barausse}, E., \& {Silk}, J. 2015, \apj, 812, 72

\bibitem[{{Antonini} {et~al.}(2012){Antonini}, {Capuzzo-Dolcetta},
  {Mastrobuono-Battisti}, \& {Merritt}}]{Antonini2012}
{Antonini}, F., {Capuzzo-Dolcetta}, R., {Mastrobuono-Battisti}, A., \&
  {Merritt}, D. 2012, \apj, 750, 111

\bibitem[{{Arca-Sedda} \& {Capuzzo-Dolcetta}(2014)}]{ArcaSedda2014}
{Arca-Sedda}, M. \& {Capuzzo-Dolcetta}, R. 2014, \apj, 785, 51

\bibitem[{{Arca-Sedda} {et~al.}(2015){Arca-Sedda}, {Capuzzo-Dolcetta},
  {Antonini}, \& {Seth}}]{ArcaSedda2015}
{Arca-Sedda}, M., {Capuzzo-Dolcetta}, R., {Antonini}, F., \& {Seth}, A. 2015,
  \apj, 806, 220

\bibitem[{{Arca Sedda} {et~al.}(2020){Arca Sedda}, {Gualandris}, {Do},
  {Feldmeier-Krause}, {Neumayer}, \& {Erkal}}]{ArcaSedda2020}
{Arca Sedda}, M., {Gualandris}, A., {Do}, T., {et~al.} 2020, \apjl, 901, L29

\bibitem[{{Arsenault} {et~al.}(2008){Arsenault}, {Madec}, {Hubin}, {Paufique},
  {Stroebele}, {Soenke}, {Donaldson}, {Fedrigo}, {Oberti}, {Tordo}, {Downing},
  {Kiekebusch}, {Conzelmann}, {Duchateau}, {Jost}, {Hackenberg}, {Bonaccini
  Calia}, {Delabre}, {Stuik}, {Biasi}, {Gallieni}, {Lazzarini}, {Lelouarn}, \&
  {Glindeman}}]{AO}
{Arsenault}, R., {Madec}, P.~Y., {Hubin}, N., {et~al.} 2008, in Society of
  Photo-Optical Instrumentation Engineers (SPIE) Conference Series, Vol. 7015,
  Adaptive Optics Systems, ed. N.~{Hubin}, C.~E. {Max}, \& P.~L. {Wizinowich},
  701524

\bibitem[{{Bacon} {et~al.}(2010){Bacon}, {Accardo}, {Adjali}, {Anwand},
  {Bauer}, {Biswas}, {Blaizot}, {Boudon}, {Brau-Nogue}, {Brinchmann},
  {Caillier}, {Capoani}, {Carollo}, {Contini}, {Couderc}, {Daguis{\'e}},
  {Deiries}, {Delabre}, {Dreizler}, {Dubois}, {Dupieux}, {Dupuy}, {Emsellem},
  {Fechner}, {Fleischmann}, {Fran{\c{c}}ois}, {Gallou}, {Gharsa}, {Glindemann},
  {Gojak}, {Guiderdoni}, {Hansali}, {Hahn}, {Jarno}, {Kelz}, {Koehler},
  {Kosmalski}, {Laurent}, {Le Floch}, {Lilly}, {Lizon}, {Loupias}, {Manescau},
  {Monstein}, {Nicklas}, {Olaya}, {Pares}, {Pasquini}, {P{\'e}contal-Rousset},
  {Pell{\'o}}, {Petit}, {Popow}, {Reiss}, {Remillieux}, {Renault}, {Roth},
  {Rupprecht}, {Serre}, {Schaye}, {Soucail}, {Steinmetz}, {Streicher}, {Stuik},
  {Valentin}, {Vernet}, {Weilbacher}, {Wisotzki}, \& {Yerle}}]{MUSE}
{Bacon}, R., {Accardo}, M., {Adjali}, L., {et~al.} 2010, in Society of
  Photo-Optical Instrumentation Engineers (SPIE) Conference Series, Vol. 7735,
  Ground-based and Airborne Instrumentation for Astronomy III, ed. I.~S.
  {McLean}, S.~K. {Ramsay}, \& H.~{Takami}, 773508

\bibitem[{{Bacon} {et~al.}(2016){Bacon}, {Piqueras}, {Conseil}, {Richard}, \&
  {Shepherd}}]{MPDAF}
{Bacon}, R., {Piqueras}, L., {Conseil}, S., {Richard}, J., \& {Shepherd}, M.
  2016, {MPDAF: MUSE Python Data Analysis Framework}, Astrophysics Source Code
  Library, record ascl:1611.003

\bibitem[{{Beasley}(2020)}]{Beasley2020}
{Beasley}, M.~A. 2020, in Reviews in Frontiers of Modern Astrophysics; From
  Space Debris to Cosmology, 245--277

\bibitem[{{Bekki}(2007)}]{Bekki2007}
{Bekki}, K. 2007, PASA, 24, 77

\bibitem[{{Bekki} {et~al.}(2006){Bekki}, {Couch}, \& {Shioya}}]{Bekki2006}
{Bekki}, K., {Couch}, W.~J., \& {Shioya}, Y. 2006, \apjl, 642, L133

\bibitem[{{Bica} {et~al.}(2020){Bica}, {Westera}, {Kerber}, {Dias}, {Maia},
  {Santos}, {Barbuy}, \& {Oliveira}}]{Bica2020}
{Bica}, E., {Westera}, P., {Kerber}, L. d.~O., {et~al.} 2020, \aj, 159, 82

\bibitem[{{B{\"o}ker} {et~al.}(2004){B{\"o}ker}, {Sarzi}, {McLaughlin}, {van
  der Marel}, {Rix}, {Ho}, \& {Shields}}]{Boker2004}
{B{\"o}ker}, T., {Sarzi}, M., {McLaughlin}, D.~E., {et~al.} 2004, \aj, 127, 105

\bibitem[{{Brodie} \& {Strader}(2006)}]{Brodie2006}
{Brodie}, J.~P. \& {Strader}, J. 2006, \araa, 44, 193

\bibitem[{{Buzzo} {et~al.}(2021){Buzzo}, {Ziegler}, {Amram}, {Verdugo},
  {Barbosa}, {Ciocan}, {Papaderos}, {Torres-Flores}, \& {Mendes de
  Oliveira}}]{Buzzo2021}
{Buzzo}, M.~L., {Ziegler}, B., {Amram}, P., {et~al.} 2021, \mnras, 503, 106

\bibitem[{{Calzetti} {et~al.}(2000){Calzetti}, {Armus}, {Bohlin}, {Kinney},
  {Koornneef}, \& {Storchi-Bergmann}}]{Calzetti2000}
{Calzetti}, D., {Armus}, L., {Bohlin}, R.~C., {et~al.} 2000, \apj, 533, 682

\bibitem[{{Cappellari}(2017)}]{Cappellari2017}
{Cappellari}, M. 2017, \mnras, 466, 798

\bibitem[{{Cappellari} \& {Copin}(2003)}]{Cappellari2003}
{Cappellari}, M. \& {Copin}, Y. 2003, \mnras, 342, 345

\bibitem[{{Cappellari} \& {Emsellem}(2004)}]{Cappellari2004}
{Cappellari}, M. \& {Emsellem}, E. 2004, \pasp, 116, 138

\bibitem[{{Capuzzo-Dolcetta}(1993)}]{CapuzzoDolcetta1993}
{Capuzzo-Dolcetta}, R. 1993, \apj, 415, 616

\bibitem[{{Capuzzo-Dolcetta} \&
  {Miocchi}(2008{\natexlab{a}})}]{CapuzzoDolcetta2008}
{Capuzzo-Dolcetta}, R. \& {Miocchi}, P. 2008{\natexlab{a}}, \apj, 681, 1136

\bibitem[{{Capuzzo-Dolcetta} \&
  {Miocchi}(2008{\natexlab{b}})}]{CapuzzoDolcetta2008a}
{Capuzzo-Dolcetta}, R. \& {Miocchi}, P. 2008{\natexlab{b}}, \mnras, 388, L69

\bibitem[{{Da Costa}(1991)}]{DaCosta1991}
{Da Costa}, G.~S. 1991, in The Magellanic Clouds, ed. R.~{Haynes} \&
  D.~{Milne}, Vol. 148, 183

\bibitem[{{Emsellem} {et~al.}(2022){Emsellem}, {Schinnerer}, {Santoro},
  {Belfiore}, {Pessa}, {McElroy}, {Blanc}, {Congiu}, {Groves}, {Ho}, {Kreckel},
  {Razza}, {Sanchez-Blazquez}, {Egorov}, {Faesi}, {Klessen}, {Leroy}, {Meidt},
  {Querejeta}, {Rosolowsky}, {Scheuermann}, {Anand}, {Barnes},
  {Be{\v{s}}li{\'c}}, {Bigiel}, {Boquien}, {Cao}, {Chevance}, {Dale},
  {Eibensteiner}, {Glover}, {Grasha}, {Henshaw}, {Hughes}, {Koch}, {Kruijssen},
  {Lee}, {Liu}, {Pan}, {Pety}, {Saito}, {Sandstrom}, {Schruba}, {Sun},
  {Thilker}, {Usero}, {Watkins}, \& {Williams}}]{Emsellem2022}
{Emsellem}, E., {Schinnerer}, E., {Santoro}, F., {et~al.} 2022, \aap, 659, A191

\bibitem[{{Erroz-Ferrer} {et~al.}(2019){Erroz-Ferrer}, {Carollo}, {den Brok},
  {Onodera}, {Brinchmann}, {Marino}, {Monreal-Ibero}, {Schaye}, {Woo},
  {Cibinel}, {Debattista}, {Inami}, {Maseda}, {Richard}, {Tacchella}, \&
  {Wisotzki}}]{ErrozFerrer2019}
{Erroz-Ferrer}, S., {Carollo}, C.~M., {den Brok}, M., {et~al.} 2019, \mnras,
  484, 5009

\bibitem[{{Fahrion} {et~al.}(2022){Fahrion}, {Leaman}, {Lyubenova}, \& {van de
  Ven}}]{Fahrion2022}
{Fahrion}, K., {Leaman}, R., {Lyubenova}, M., \& {van de Ven}, G. 2022, \aap,
  658, A172

\bibitem[{{Fahrion} {et~al.}(2020{\natexlab{a}}){Fahrion}, {Lyubenova},
  {Hilker}, {van de Ven}, {Falc{\'o}n-Barroso}, {Leaman},
  {Mart{\'\i}n-Navarro}, {Bittner}, {Coccato}, {Corsini}, {Gadotti}, {Iodice},
  {McDermid}, {Pinna}, {Sarzi}, {Viaene}, {de Zeeuw}, \& {Zhu}}]{Fahrion2020c}
{Fahrion}, K., {Lyubenova}, M., {Hilker}, M., {et~al.} 2020{\natexlab{a}},
  \aap, 637, A27

\bibitem[{{Fahrion} {et~al.}(2021){Fahrion}, {Lyubenova}, {van de Ven},
  {Hilker}, {Leaman}, {Falc{\'o}n-Barroso}, {Bittner}, {Coccato}, {Corsini},
  {Gadotti}, {Iodice}, {McDermid}, {Mart{\'\i}n-Navarro}, {Pinna}, {Poci},
  {Sarzi}, {de Zeeuw}, \& {Zhu}}]{Fahrion2021}
{Fahrion}, K., {Lyubenova}, M., {van de Ven}, G., {et~al.} 2021, \aap, 650,
  A137

\bibitem[{{Fahrion} {et~al.}(2019){Fahrion}, {Lyubenova}, {van de Ven},
  {Leaman}, {Hilker}, {Mart{\'\i}n-Navarro}, {Zhu}, {Alfaro-Cuello}, {Coccato},
  {Corsini}, {Falc{\'o}n-Barroso}, {Iodice}, {McDermid}, {Sarzi}, \& {de
  Zeeuw}}]{Fahrion2019}
{Fahrion}, K., {Lyubenova}, M., {van de Ven}, G., {et~al.} 2019, \aap, 628, A92

\bibitem[{{Fahrion} {et~al.}(2020{\natexlab{b}}){Fahrion}, {M{\"u}ller},
  {Rejkuba}, {Hilker}, {Lyubenova}, {van de Ven}, {Georgiev}, {Lelli},
  {Pawlowski}, \& {Jerjen}}]{Fahrion2020}
{Fahrion}, K., {M{\"u}ller}, O., {Rejkuba}, M., {et~al.} 2020{\natexlab{b}},
  \aap, 634, A53

\bibitem[{{Falc{\'o}n-Barroso} {et~al.}(2011){Falc{\'o}n-Barroso}, {van de
  Ven}, {Peletier}, {Bureau}, {Jeong}, {Bacon}, {Cappellari}, {Davies}, {de
  Zeeuw}, {Emsellem}, {Krajnovi{\'c}}, {Kuntschner}, {McDermid}, {Sarzi},
  {Shapiro}, {van den Bosch}, {van der Wolk}, {Weijmans}, \&
  {Yi}}]{FalconBarroso2011}
{Falc{\'o}n-Barroso}, J., {van de Ven}, G., {Peletier}, R.~F., {et~al.} 2011,
  \mnras, 417, 1787

\bibitem[{{Feldmeier-Krause} {et~al.}(2020){Feldmeier-Krause}, {Kerzendorf},
  {Do}, {Nogueras-Lara}, {Neumayer}, {Walcher}, {Seth}, {Sch{\"o}del}, {de
  Zeeuw}, {Hilker}, {L{\"u}tzgendorf}, {Kuntschner}, \&
  {Kissler-Patig}}]{FeldmeierKrause2020}
{Feldmeier-Krause}, A., {Kerzendorf}, W., {Do}, T., {et~al.} 2020, \mnras, 494,
  396

\bibitem[{{Feldmeier-Krause} {et~al.}(2015){Feldmeier-Krause}, {Neumayer},
  {Sch{\"o}del}, {Seth}, {Hilker}, {de Zeeuw}, {Kuntschner}, {Walcher},
  {L{\"u}tzgendorf}, \& {Kissler-Patig}}]{FeldmeierKrause2015}
{Feldmeier-Krause}, A., {Neumayer}, N., {Sch{\"o}del}, R., {et~al.} 2015, \aap,
  584, A2

\bibitem[{{Fensch} {et~al.}(2014){Fensch}, {Mieske}, {M{\"u}ller-Seidlitz}, \&
  {Hilker}}]{Fensch2014}
{Fensch}, J., {Mieske}, S., {M{\"u}ller-Seidlitz}, J., \& {Hilker}, M. 2014,
  \aap, 567, A105

\bibitem[{{Geisler} {et~al.}(1997){Geisler}, {Bica}, {Dottori}, {Claria},
  {Piatti}, \& {Santos}}]{Geisler1997}
{Geisler}, D., {Bica}, E., {Dottori}, H., {et~al.} 1997, \aj, 114, 1920

\bibitem[{{Georgiev} \& {B{\"o}ker}(2014)}]{Georgiev2014}
{Georgiev}, I.~Y. \& {B{\"o}ker}, T. 2014, \mnras, 441, 3570

\bibitem[{{Georgiev} {et~al.}(2016){Georgiev}, {B{\"o}ker}, {Leigh},
  {L{\"u}tzgendorf}, \& {Neumayer}}]{Georgiev2016}
{Georgiev}, I.~Y., {B{\"o}ker}, T., {Leigh}, N., {L{\"u}tzgendorf}, N., \&
  {Neumayer}, N. 2016, \mnras, 457, 2122

\bibitem[{{Georgiev} {et~al.}(2009{\natexlab{a}}){Georgiev}, {Hilker}, {Puzia},
  {Goudfrooij}, \& {Baumgardt}}]{Georgiev2009b}
{Georgiev}, I.~Y., {Hilker}, M., {Puzia}, T.~H., {Goudfrooij}, P., \&
  {Baumgardt}, H. 2009{\natexlab{a}}, \mnras, 396, 1075

\bibitem[{{Georgiev} {et~al.}(2009{\natexlab{b}}){Georgiev}, {Puzia}, {Hilker},
  \& {Goudfrooij}}]{Georgiev2009a}
{Georgiev}, I.~Y., {Puzia}, T.~H., {Hilker}, M., \& {Goudfrooij}, P.
  2009{\natexlab{b}}, \mnras, 392, 879

\bibitem[{Gnedin {et~al.}(2014)Gnedin, Ostriker, \& Tremaine}]{Gnedin2014}
Gnedin, O.~Y., Ostriker, J.~P., \& Tremaine, S. 2014, The Astrophysical
  Journal, 785, 71

\bibitem[{{Gu{\'e}rou} {et~al.}(2016){Gu{\'e}rou}, {Emsellem}, {Krajnovi{\'c}},
  {McDermid}, {Contini}, \& {Weilbacher}}]{Guerou2016}
{Gu{\'e}rou}, A., {Emsellem}, E., {Krajnovi{\'c}}, D., {et~al.} 2016, \aap,
  591, A143

\bibitem[{{Guillard} {et~al.}(2016){Guillard}, {Emsellem}, \&
  {Renaud}}]{Guillard2016}
{Guillard}, N., {Emsellem}, E., \& {Renaud}, F. 2016, \mnras, 461, 3620

\bibitem[{{Hannah} {et~al.}(2021){Hannah}, {Seth}, {Nguyen}, {Dumont},
  {Kacharov}, {Neumayer}, \& {den Brok}}]{Hannah2021}
{Hannah}, C.~H., {Seth}, A.~C., {Nguyen}, D.~D., {et~al.} 2021, \aj, 162, 281

\bibitem[{{Hao} {et~al.}(2011){Hao}, {Kennicutt}, {Johnson}, {Calzetti},
  {Dale}, \& {Moustakas}}]{Hao2011}
{Hao}, C.-N., {Kennicutt}, R.~C., {Johnson}, B.~D., {et~al.} 2011, \apj, 741,
  124

\bibitem[{{Johnston} {et~al.}(2020){Johnston}, {Puzia}, {D'Ago}, {Eigenthaler},
  {Galaz}, {H{\"a}u{\ss}ler}, {Mora}, {Ordenes-Brice{\~n}o}, {Rong},
  {Spengler}, {Vogt}, {C{\^o}t{\'e}}, {Grebel}, {Hilker}, {Mieske}, {Miller},
  {S{\'a}nchez-Janssen}, {Taylor}, \& {Zhang}}]{Johnston2020}
{Johnston}, E.~J., {Puzia}, T.~H., {D'Ago}, G., {et~al.} 2020, \mnras, 495,
  2247

\bibitem[{{Kacharov} {et~al.}(2018){Kacharov}, {Neumayer}, {Seth},
  {Cappellari}, {McDermid}, {Walcher}, \& {B{\"o}ker}}]{Kacharov2018}
{Kacharov}, N., {Neumayer}, N., {Seth}, A.~C., {et~al.} 2018, \mnras, 480, 1973

\bibitem[{{Kroupa} \& {Weidner}(2003)}]{Kroupa2003}
{Kroupa}, P. \& {Weidner}, C. 2003, \apj, 598, 1076

\bibitem[{{Lamers} {et~al.}(2017){Lamers}, {Kruijssen}, {Bastian}, {Rejkuba},
  {Hilker}, \& {Kissler-Patig}}]{Lamers2017}
{Lamers}, H.~J.~G.~L.~M., {Kruijssen}, J.~M.~D., {Bastian}, N., {et~al.} 2017,
  \aap, 606, A85

\bibitem[{{Leaman} {et~al.}(2020){Leaman}, {Ruiz-Lara}, {Cole}, {Beasley},
  {Boecker}, {Fahrion}, {Bianchini}, {Falc{\'o}n-Barroso}, {Webb}, {Sills},
  {Mastrobuono-Battisti}, {Neumayer}, \& {Sippel}}]{Leaman2020}
{Leaman}, R., {Ruiz-Lara}, T., {Cole}, A.~A., {et~al.} 2020, \mnras, 492, 5102

\bibitem[{{Leaman} {et~al.}(2013){Leaman}, {VandenBerg}, \&
  {Mendel}}]{Leaman2013}
{Leaman}, R., {VandenBerg}, D.~A., \& {Mendel}, J.~T. 2013, \mnras, 436, 122

\bibitem[{{Loose} {et~al.}(1982){Loose}, {Kruegel}, \& {Tutukov}}]{Loose1982}
{Loose}, H.~H., {Kruegel}, E., \& {Tutukov}, A. 1982, \aap, 105, 342

\bibitem[{{Lyubenova} \& {Tsatsi}(2019)}]{Lyubenova2019}
{Lyubenova}, M. \& {Tsatsi}, A. 2019, \aap, 629, A44

\bibitem[{{Martocchia} {et~al.}(2018){Martocchia}, {Cabrera-Ziri}, {Lardo},
  {Dalessandro}, {Bastian}, {Kozhurina-Platais}, {Usher}, {Niederhofer},
  {Cordero}, {Geisler}, {Hollyhead}, {Kacharov}, {Larsen}, {Li}, {Mackey},
  {Hilker}, {Mucciarelli}, {Platais}, \& {Salaris}}]{Martocchia2018}
{Martocchia}, S., {Cabrera-Ziri}, I., {Lardo}, C., {et~al.} 2018, \mnras, 473,
  2688

\bibitem[{{Milosavljevi{\'c}}(2004)}]{Milosavljevic2004}
{Milosavljevi{\'c}}, M. 2004, \apj, 605, L13

\bibitem[{{Neumayer} {et~al.}(2020){Neumayer}, {Seth}, \&
  {B{\"o}ker}}]{Neumayer2020}
{Neumayer}, N., {Seth}, A., \& {B{\"o}ker}, T. 2020, \aapr, 28, 4

\bibitem[{{Nguyen} {et~al.}(2014){Nguyen}, {Seth}, {Reines}, {den Brok},
  {Sand}, \& {McLeod}}]{Nguyen2014}
{Nguyen}, D.~D., {Seth}, A.~C., {Reines}, A.~E., {et~al.} 2014, \apj, 794, 34

\bibitem[{{Osterbrock} \& {Ferland}(2006)}]{OsterbrockFerland2006}
{Osterbrock}, D.~E. \& {Ferland}, G.~J. 2006, {Astrophysics of gaseous nebulae
  and active galactic nuclei}

\bibitem[{{Paudel} \& {Yoon}(2020)}]{Paudel2020}
{Paudel}, S. \& {Yoon}, S.-J. 2020, \apjl, 898, L47

\bibitem[{{Peng} {et~al.}(2006){Peng}, {Jord{\'a}n}, {C{\^o}t{\'e}},
  {Blakeslee}, {Ferrarese}, {Mei}, {West}, {Merritt}, {Milosavljevi{\'c}}, \&
  {Tonry}}]{Peng2006}
{Peng}, E.~W., {Jord{\'a}n}, A., {C{\^o}t{\'e}}, P., {et~al.} 2006, \apj, 639,
  95

\bibitem[{{Pietrinferni} {et~al.}(2004){Pietrinferni}, {Cassisi}, {Salaris}, \&
  {Castelli}}]{Pietrinferni2004}
{Pietrinferni}, A., {Cassisi}, S., {Salaris}, M., \& {Castelli}, F. 2004, \apj,
  612, 168

\bibitem[{{Pietrinferni} {et~al.}(2006){Pietrinferni}, {Cassisi}, {Salaris}, \&
  {Castelli}}]{Pietrinferni2006}
{Pietrinferni}, A., {Cassisi}, S., {Salaris}, M., \& {Castelli}, F. 2006, \apj,
  642, 797

\bibitem[{{Pinna} {et~al.}(2021){Pinna}, {Neumayer}, {Seth}, {Emsellem},
  {Nguyen}, {B{\"o}ker}, {Cappellari}, {McDermid}, {Voggel}, \&
  {Walcher}}]{Pinna2021}
{Pinna}, F., {Neumayer}, N., {Seth}, A., {et~al.} 2021, \apj, 921, 8

\bibitem[{{Portaluri} {et~al.}(2013){Portaluri}, {Corsini}, {Morelli},
  {Hartmann}, {Dalla Bont{\`a}}, {Debattista}, \& {Pizzella}}]{Portaluri2013}
{Portaluri}, E., {Corsini}, E.~M., {Morelli}, L., {et~al.} 2013, \mnras, 433,
  434

\bibitem[{{Puzia} \& {Sharina}(2008)}]{Puzia2008}
{Puzia}, T.~H. \& {Sharina}, M.~E. 2008, \apj, 674, 909

\bibitem[{{S{\'a}nchez-Janssen} {et~al.}(2019){S{\'a}nchez-Janssen},
  {C{\^o}t{\'e}}, {Ferrarese}, {Peng}, {Roediger}, {Blakeslee}, {Emsellem},
  {Puzia}, {Spengler}, {Taylor}, {{\'A}lamo-Mart{\'\i}nez}, {Boselli},
  {Cantiello}, {Cuillandre}, {Duc}, {Durrell}, {Gwyn}, {MacArthur},
  {Lan{\c{c}}on}, {Lim}, {Liu}, {Mei}, {Miller}, {Mu{\~n}oz}, {Mihos},
  {Paudel}, {Powalka}, \& {Toloba}}]{SanchezJanssen2019}
{S{\'a}nchez-Janssen}, R., {C{\^o}t{\'e}}, P., {Ferrarese}, L., {et~al.} 2019,
  \apj, 878, 18

\bibitem[{{Schinnerer} {et~al.}(2007){Schinnerer}, {B{\"o}ker}, {Emsellem}, \&
  {Downes}}]{Schinnerer2007}
{Schinnerer}, E., {B{\"o}ker}, T., {Emsellem}, E., \& {Downes}, D. 2007, \aap,
  462, L27

\bibitem[{{Seth} {et~al.}(2010){Seth}, {Cappellari}, {Neumayer}, {Caldwell},
  {Bastian}, {Olsen}, {Blum}, {Debattista}, {McDermid}, {Puzia}, \&
  {Stephens}}]{Seth2010}
{Seth}, A.~C., {Cappellari}, M., {Neumayer}, N., {et~al.} 2010, \apj, 714, 713

\bibitem[{{Str{\"o}bele} {et~al.}(2012){Str{\"o}bele}, {La Penna}, {Arsenault},
  {Conzelmann}, {Delabre}, {Duchateau}, {Dorn}, {Fedrigo}, {Hubin}, {Quentin},
  {Jolley}, {Kiekebusch}, {Kirchbauer}, {Klein}, {Kolb}, {Kuntschner}, {Le
  Louarn}, {Lizon}, {Madec}, {Pettazzi}, {Soenke}, {Tordo}, {Vernet}, \&
  {Muradore}}]{Galacsi}
{Str{\"o}bele}, S., {La Penna}, P., {Arsenault}, R., {et~al.} 2012, in Society
  of Photo-Optical Instrumentation Engineers (SPIE) Conference Series, Vol.
  8447, Adaptive Optics Systems III, ed. B.~L. {Ellerbroek}, E.~{Marchetti}, \&
  J.-P. {V{\'e}ran}, 844737

\bibitem[{{Su} {et~al.}(2022){Su}, {Salo}, {Janz}, {Venhola}, \&
  {Peletier}}]{Su2022}
{Su}, A.~H., {Salo}, H., {Janz}, J., {Venhola}, A., \& {Peletier}, R.~F. 2022,
  arXiv e-prints, arXiv:2206.01490

\bibitem[{{Tremaine} {et~al.}(1975){Tremaine}, {Ostriker}, \&
  {Spitzer}}]{Tremaine1975}
{Tremaine}, S.~D., {Ostriker}, J.~P., \& {Spitzer}, Jr., L. 1975, \apj, 196,
  407

\bibitem[{{Turner} {et~al.}(2012){Turner}, {C{\^o}t{\'e}}, {Ferrarese},
  {Jord{\'a}n}, {Blakeslee}, {Mei}, {Peng}, \& {West}}]{Turner2012}
{Turner}, M.~L., {C{\^o}t{\'e}}, P., {Ferrarese}, L., {et~al.} 2012, \apjs,
  203, 5

\bibitem[{{Vazdekis} {et~al.}(1996){Vazdekis}, {Casuso}, {Peletier}, \&
  {Beckman}}]{Vazdekis1996}
{Vazdekis}, A., {Casuso}, E., {Peletier}, R.~F., \& {Beckman}, J.~E. 1996,
  \apjs, 106, 307

\bibitem[{{Vazdekis} {et~al.}(2016){Vazdekis}, {Koleva}, {Ricciardelli},
  {R{\"o}ck}, \& {Falc{\'o}n-Barroso}}]{Vazdekis2016}
{Vazdekis}, A., {Koleva}, M., {Ricciardelli}, E., {R{\"o}ck}, B., \&
  {Falc{\'o}n-Barroso}, J. 2016, \mnras, 463, 3409

\bibitem[{{Vazdekis} {et~al.}(2010){Vazdekis}, {S{\'a}nchez-Bl{\'a}zquez},
  {Falc{\'o}n-Barroso}, {Cenarro}, {Beasley}, {Cardiel}, {Gorgas}, \&
  {Peletier}}]{Vazdekis2010}
{Vazdekis}, A., {S{\'a}nchez-Bl{\'a}zquez}, P., {Falc{\'o}n-Barroso}, J.,
  {et~al.} 2010, \mnras, 404, 1639

\bibitem[{{Weilbacher} {et~al.}(2020){Weilbacher}, {Palsa}, {Streicher},
  {Bacon}, {Urrutia}, {Wisotzki}, {Conseil}, {Husemann}, {Jarno}, {Kelz},
  {P{\'e}contal-Rousset}, {Richard}, {Roth}, {Selman}, \&
  {Vernet}}]{Weilbacher2020}
{Weilbacher}, P.~M., {Palsa}, R., {Streicher}, O., {et~al.} 2020, \aap, 641,
  A28

\bibitem[{{Yu} {et~al.}(2020){Yu}, {Ho}, \& {Wang}}]{Yu2020}
{Yu}, N., {Ho}, L.~C., \& {Wang}, J. 2020, \apj, 898, 102

\bibitem[{{Zhang} {et~al.}(2018){Zhang}, {Puzia}, {Peng}, {Liu},
  {C{\^o}t{\'e}}, {Ferrarese}, {Duc}, {Eigenthaler}, {Lim}, {Lan{\c{c}}on},
  {Mu{\~n}oz}, {Roediger}, {S{\'a}nchez-Janssen}, {Taylor}, \&
  {Yu}}]{Zhang2018}
{Zhang}, H.-X., {Puzia}, T.~H., {Peng}, E.~W., {et~al.} 2018, \apj, 858, 37

\end{thebibliography}

\end{document}